\documentclass{aa}
\usepackage{txfonts}
\usepackage{graphicx}
\usepackage{natbib}
\usepackage[ruled,vlined]{algorithm2e}
\bibpunct{(}{)}{;}{a}{}{,}

\begin{document}

\title{How accurate are the time delay estimates \\ in gravitational lensing?}
\author{Juan C. Cuevas-Tello \inst{1,3} \and Peter Ti\v{n}o\inst{1} \and Somak Raychaudhury\inst{2}}
\institute{School of Computer Science, University of Birmingham, Edgbaston, Birmingham B15 2TT, United Kingdom \and
School of Physics and Astronomy, University of Birmingham, Edgbaston, Birmingham B15 2TT, United Kingdom \and
Engineering Faculty, Autonomous University of San Luis Potos\'{i}, M\'{e}xico}

\offprints{somak@star.sr.bham.ac.uk}

\date{Received in December 2005 / Accepted in April 2006}
\bibliographystyle {aa}

\abstract{We present a novel approach to estimate the time delay
between light curves of multiple images in a gravitationally lensed
system, based on Kernel methods in the context of machine learning. 
We perform various experiments with artificially
generated irregularly-sampled data sets to study the effect of the
various levels of noise and the presence of gaps of various size in
the monitoring data. We compare the performance of our method with
various other popular methods of estimating the time delay and
conclude, from experiments with artificial data, that our method is
least vulnerable to missing data and irregular sampling, within
reasonable bounds of Gaussian noise.  Thereafter, we use our method to
determine the time delays between the two images of quasar \object{Q0957+561}
from radio monitoring data at 4~cm and 6~cm, and conclude that if only
the observations at epochs common to both wavelengths are used, the
time delay gives consistent estimates, which can be combined to yield
$408\pm 12$ days.  The full 6~cm dataset, which covers a longer
monitoring period, yields a value which is 10\% larger, but this can
be attributed to differences in sampling and missing data.
 }

%\titlerunning{How accurate are the time delay estimates?}
\authorrunning{Cuevas-Tello, J.C. et al.}

\maketitle

\section{Introduction}

Long before the first gravitationally lensed quasar was discovered in
1979 \citep{Walsh:1979}, Refsdal suggested that time delays of source
fluctuations between the multiple images could be used to measure the
Universe \citep{Refsdal:1964:R64, Refsdal:1966:R66}.  This first
lensed quasar, \object{Q0957+561}, is also the most studied so far (Fig.
\ref{0957-fig}), and many attempts have been made to estimate the time
delay between its two principal images.

The measurement of the delay between the images A and B of \object{Q0957+561}
has been the subject of sensitive controversy ever since the first
claim of measurement in the early 1980s.
\citet{Haarsma:1997:T6CM} reviews the various
measurements, showing how various delays in the range of 300 to 1000 days have
been claimed, from various data sets using different methods
\citep{Kochanek:2004:THC}.  
In the early nineties, the quoted time delay values were either around
420~days \citep[e.g.,][]{Falco:1991:GL} or 540~days
\citep[e.g.,][]{Press:1992:TTD},
culminating in a ``definitive'' measure of a time delay of 417$\pm$3 days
\citep{Kundic:1997:ARD}.  

More recently, efforts have concentrated on characterising the errors
on such measurements \citep[e.g.,][]{Pindor:2005:DGL}.  The case of
\object{Q0957+561} also illustrates this.  Variously, from optical
monitoring data,
\citet{Ovaldsen:2003:NAP}, \citet{Oscoz:2001:TDI},
\citet{Burud:2001:ANA} and \citet{Colley:2003:ATC} 
estimate a time delay of 424.9$\pm$1.2 422.6$\pm$0.6, 423$\pm$9 and
417$\pm$0.07 days respectively (more estimates are in
Table~\ref{review_time_delay}).  Given the error bars, many of these measures
are inconsistent with the definitive value quoted above, and often
with each other, if taken at face value.

To measure the time delay between signals arriving from the same
source but via different paths, typically ranging from a few days to a
few years in the data available so far, one needs to frequently
observe the same set of sources over long periods of time. Due to the
usual methods of allocation of telescope time and the natural
time scale of projects, the data obtained typically are not regularly
sampled, and could be obtained by a wide range of instruments and at
different frequencies, often with large gaps in the time series.
Since the use of time delays in constraining cosmological parameters
\citep[e.g.,][]{Saha:2004:TDQ} requires these delays to
be measured to a precision and reliability that is better than
afforded by current practice, it is important to look for better and
more robust methods, where the dependence of the results on the
incompleteness of the data is well understood. 

Rigorous error estimates based on a functional form for random errors
often well represent the inherent limitations of a particular method,
but do not address systematic effects due to sampling strategies, to 
the heterogeneity of monitoring programmes  and those due to
data missing for various practical reasons, like observing schedules.
With the bank of data for light curves of lens systems growing rapidly, 
the effect
of such systematics on the measured values 
of time delay need to be well understood.
This is particularly
important in view of future missions like the Large Synoptic Survey
Telescope (LSST) and Supernova/Acceleration Probe (SNAP), which will
make large monitoring data sets available for hundreds of
multiply-imaged distant sources,
\citep[e.g.,][]{Mortsell:2005:GL,Fassnacht:2004:GL}, thus
rendering them major statistical tools for cosmological purposes.

We present a novel approach to the problem of determining the delay
between noisy time-dependent signals that have been measured at
irregular intervals over several years, often with large gaps in the
monitoring programme.  Ours is an automatic method that allows us to
analyse large-scale experiments more accurately than typical
methods. We study the effect of gaps of various length, regularly or
irregularly sampled, in the monitoring data, in addition to the
different levels of noise. This study should provide some insight for
astronomers designing future observational campaigns for monitoring
multiple-images quasars.

As an illustration of the application of our method, we apply it here
to radio observations, at 4~cm and 6~cm, of the well known
gravitational lens \object{Q0957+561}
\citep{Haarsma:1999:TRW}, and compare the results to other studies using
the same or similar datasets.

The remainder of this article is organised as follows: in
\S\ref{kernel-method-section}, we present our
method. \S\ref{methods-section} is a survey of methods to estimate the
time delay presenting a detailed review of three of the most popular
methods.  \S\ref{artificial-section} describes the artificial data
generated to perform our simulations.
\S\ref{results-artificial-section} shows the results on these
artificial data. In
\S\ref{0957-section} we present estimates 
for the time delay between the two principal
images of \object{Q0957+561} from radio data at 4~cm and 6~cm, using
our methods presented here,
followed by a concluding summary.

\section{The model}\label{kernel-method-section}

We model the observed flux at a given frequency (in the radio
or optical range) from two lensed images A and B
of the same distant source,
as two time series
\begin{equation}
  \label{data}
      x_{A}(t_{i})=h_{A}(t_{i})+\varepsilon_{A}(t_{i}) \\
      x_{B}(t_{i})=M \cdot h_{B}(t_{i})+\varepsilon_{B}(t_{i}),
 \end{equation}
\noindent where $M$ is the ratio
of the fluxes of the two images, and $t_{i},i=1,2,...,n$ are
discrete observation times. The observation errors
$\varepsilon_{A}(t_{i})$ and
$\varepsilon_{B}(t_{i})$ are
modelled as zero-mean Normal distributions
\begin{equation}
  \label{error-model}
   N(0,\sigma_{A}(t_{i})) \ \ \hbox{and} \ \ N(0,\sigma_{B}(t_{i})),
\end{equation}
\noindent respectively. Now,
\begin{equation}
  \label{hA}
    h_{A}(t_{i})=\sum_{j=1}^N\alpha_{j}K(c_{j},t_{i})
\end{equation}
\noindent is the ``underlying'' light curve that underpins
image~A, whereas
\begin{equation}
  \label{hB}
    h_{B}(t_{i})=\sum_{j=1}^N\alpha_{j}K(c_{j}+\Delta,t_{i})
\end{equation}
is a time-delayed (by $\Delta$) version of $h_{A}(t_{i})$
underpinning image~B.

The functions $h_{A}$ and $h_B$ are formulated within the generalised
linear regression framework \citep[e.g.][\S2]{Shawe-Taylor:2004:KM}. Each function is a linear superposition of
$N$ kernels $K(\cdot,\cdot)$ centred at either $c_j$, $j=1,2,...,N$
(function $f_A$), or $c_j+\Delta$, $j=1,2,...,N$ (function $f_B$).
The model (\ref{data})-(\ref{hB}) has $N$ free parameters
$\alpha_{j}$, $j=1,2,...,N$, that need to be determined by
(learned from) the data.
We use Gaussian kernels of width $\omega^{2}$: \
for $c,t \in \Re$,
\begin{equation}
  \label{kernel}
    K(c,t)=\exp\, {\frac{-|t-c|^{2}}{\omega_c^{2}}}.
    \end{equation}
The kernel width $\omega_c>0$ determines the `degree of smoothness'
of the underlying curves $h_A$ and $h_B$. We describe setting of
$\omega_j = \omega_{c_j}$ and regression weights $\alpha_j$
in the next subsections.
In this study, we position kernels on all observations,
i.e. $N=n$.

Finally, our aim is to estimate the time delay $\Delta$ between
the temporal light curves corresponding
to images A and B. Given the observed data,
the likelihood of our model reads
\begin{equation}
  \label{pdata}
      P({\rm Data}\ | \ {\rm Model}) = \prod_{i=1}^{n}
      p(x_A(t_i),x_B(t_i)\ |\ \Delta,\{\alpha_{j}\}),
\end{equation}
where
\begin{eqnarray}
p(x_A(t_i),x_B(t_i)\ |\ \Delta,\{\alpha_{j}\})
& = &
\frac{1}{2 \pi \sigma_{A}^{2}(t_{i}) \sigma_{B}^{2}(t_{i})}
\nonumber \\
& &
\exp \left\{ \frac{(x_{A}(t_{i})-h_{A}(t_{i}))^{2} }{2 \sigma_{A}^{2}(t_{i})}
\right\}
\nonumber \\
& &
\exp \left\{ \frac{(x_{B}(t_{i})-M \cdot h_{B}(t_{i}))^{2}}{2 \sigma_{B}^{2}(t_{i})}
\right\}.
\end{eqnarray}

The negative log-likelihood (without constant terms)
  simplifies to
\begin{equation}
  \label{log-likelihood}
   Q = \sum_{i=1}^{n} \left( \frac{(x_{A}(t_{i})-h_{A}(t_{i}))^{2}}{\sigma_{A}^{2}(t_{i})}+
                         \frac{(x_{B}(t_{i})-M \cdot h_{B}(t_{i}))^{2}}{\sigma_{B}^{2}(t_{i})} \right).
\end{equation}

To avoid extrapolation when we apply a time delay to our underlying
curve, we do not evaluate the goodness of fit over all
observations:
\begin{equation}
  \label{fit}
   Q = \sum_{u=1}^{n-b_{1}}\frac{(x_{A}(t_{u})-h_{A}(t_{u}))^{2}}{\sigma_{A}^{2}(t_{u})}+
       \sum_{v=b_{2}}^{n}\frac{(x_{B}(t_{v})-M \cdot h_{B}(t_{v}))^{2}}{\sigma_{B}^{2}(t_{v})},
\end{equation}
\noindent
where $b_{1}$ is the greatest index
satisfying $t_{n-b_{1}} \leq t_{n}-\Delta_{max}$, and $b_{2}$ is
the smallest index
satisfying $t_{b_{2}} \geq t_{1}+\Delta_{max}$. Here,
$\Delta_{max}$ is the maximum possible time
delay we are willing to consider (fixed).

We determine
the model parameters and evaluate Eq.~(\ref{fit}) for a series of
trial values of $\Delta$. The time delay is then estimated
as the value of $\Delta$ with minimal cost (\ref{fit}).
Note that if
the errors cannot be modelled as Gaussian,
Eq.~(\ref{fit}) would need to be rewritten using an appropriate
noise term.

\subsection{Weights}\label{weights-section}

We rewrite Eq.~(\ref{log-likelihood}) as
\begin{equation}
  \label{log-likelihood-v2}
   Q = \sum_{i=1}^{n} \left( \left[ \frac{x_{A}(t_{i})}{\sigma_{A}(t_{i})}-\frac{h_{A}(t_{i})}{\sigma_{A}(t_{i})} \right]^{2} +
                       \left[  \frac{x_{B}(t_{i})}{\sigma_{B}(t_{i})}- \frac{M \cdot h_{B}(t_{i})}{\sigma_{B}(t_{i})} \right]^{2} \right).
\end{equation}
\noindent
Since we expect each of the two terms in (\ref{log-likelihood-v2})
to be individually equal to zero, we impose
\begin{equation}
  \label{alphas}
   \vec{K} \vec{\alpha} = \vec{x},
\end{equation}
\noindent where
$\vec{\alpha} = (\alpha_1, \alpha_2, ..., \alpha_N)^T$,
\begin{equation}
\label{alpha-kernel_data}
  \vec{K} =  \left[
              \begin{array}{ccc}
                K_{A}(c_{1},t_{1}) & \cdots & K_{A}(c_{N},t_{1}) \\
                \vdots         & \ddots & \vdots         \\
                K_{A}(c_{1},t_{n}) & \cdots & K_{A}(c_{N},t_{n}) \\ \hline
                K_{B}(c_{1},t_{1}) & \cdots & K_{B}(c_{N},t_{1}) \\
                \vdots         & \ddots & \vdots         \\
                K_{B}(c_{1},t_{n}) & \cdots & K_{B}(c_{N},t_{n})
              \end{array}
             \right],
\  \ \ \ \ \ \ \
\vec{x} =
            \left[
              \begin{array}{c}
                \frac{x_{A}(t_{1})}{\sigma_{A}(t_{1})} \\
                \vdots \\
                \frac{x_{A}(t_{n})}{\sigma_{A}(t_{n})} \\ \hline
                \frac{x_{B}(t_{1})}{\sigma_{B}(t_{1})} \\
                \vdots \\
                \frac{x_{B}(t_{n})}{\sigma_{B}(t_{n})}
              \end{array}
             \right],
\end{equation}
and the kernels $K_A(\cdot,\cdot)$, $K_B(\cdot,\cdot)$ have the form:
\begin{equation}
  \label{kA-kB}
    K_{A}(c,t) = \frac{K(c,t)}{\sigma_{A}(t)}, \ \ \ \ \ \ \
    K_{B}(c,t) = \frac{M \cdot K(c+\Delta,t)}{\sigma_{B}(t)}.
\end{equation}

Hence,
\begin{equation}
  \label{inverse}
   \vec{\alpha} = \vec{K}^{+}\vec{x}.
\end{equation}
We regularise the inversion
in (\ref{inverse}) through
singular value decomposition (SVD).

\subsection{Kernel parameters}\label{centres-section}

In general, in order to use Gaussian kernels (\ref{kernel})
in generalised linear regression (\ref{data})-(\ref{hB}),
the kernel positions
$c_j$, as well as kernel widths
$\omega_j$, need to be determined \citep[\S9]{Shawe-Taylor:2004:KM}.
Several approaches have been taken in the literature.
For instance, those who use
radial basis function (RBF) networks employ
e.g. $k$-means clustering, or EM algorithm and
Gaussian mixture modelling \citep[e.g. see][]{Haykin:1999:book,Hastie:2001:book}\footnote{Some approaches
attempt to simultaneously optimise the number of kernels.}.
We have explored two
approaches to kernel positioning: {\bf (i)} the centres $c_j$ uniformly 
distributed across the input range
and {\bf (ii)} the centres $c_j$ positioned at input samples $t_j$,
$j=1,2,...,n$.
The latter approach lead to superior performance and the results reported
in this paper were obtained using kernels centred at observation times
$t_j$.
As for the kernel widths,
we propose two approaches: {\bf (j)} fixed width $\omega$ and
{\bf (jj)} variable widths $\omega_j$,
$j=1,2,...,n$.
Both are described in the following
subsections.

\subsubsection{Fixed kernel width}\label{CV-section}

The width of the kernels determines the degree of smoothing for the
underlying flux curves (\ref{hA}) and (\ref{hB}).  Finding
`appropriate' values of smoothing parameters is one of the challenges
in non- and semi-parametric regression.  We use cross validation
\citep[\S 7.10]{Hastie:2001:book} to find the `optimal' kernel width
$\omega$.  In particular, we invoke a variant of
five-fold-cross-validation. We start by dividing the data set uniformly
into five blocks.  In the first step, we construct a validation set as
a collection of the first elements of each block. The validation set
has five elements. The training set is formed by the remaining
observations, i.e.  the observations not included in the validation
set. We fit our models on the training set and determine the mean
square error (MSE) over a range of delay values $\Delta$ on the
validation set.  In the next step, we construct a new validation set
as a collection of the second elements of each block. The new training
set is again formed by the remaining observations.  As before we fit
our models on the training set and determine MSE on the validation
set.  We repeat this procedure $r$ times, where $r$ is the number of
observations in each block.  Finally, the mean of all such mean square
errors (there is $r$ of them), MSE$_{CV}$, is calculated.  The kernel
width $\omega$ selected using the cross-validation is the kernel width
yielding the smallest MSE$_{CV}$.  The scheme is summarised in
Algorithm \ref{cross-validation}.

\begin{algorithm}
\caption{Cross validation\label{IR}}
\dontprintsemicolon
\label{cross-validation}
 Fix $M$, $LowerBound$ and $UpperBound$ \;
 Fix $Blocks \leftarrow 5$ \;
 Fix $PointsPerBlock \leftarrow min(\{b_{1},n-b_{2}\})/Blocks$ \;
 \For{$omega \leftarrow LowerBound$  \hbox{\rm \bf to}  $UpperBound$}
  {
   \For{$i \leftarrow 1$  \hbox{\rm \bf to} $PointsPerBlock$}
      {
      Remove the $i^{th}$ observation of each block and include it
      in the validation set\;
      \For{$\Delta \leftarrow \Delta_{min}$ \hbox{\rm \bf to} $\Delta_{max}$}
      {
        Get weights $\vec{\alpha}$ on the training set
        (using eq. (\ref{inverse}))\;
        Compute $h_{A}(t_{u})$ and $h_{B}(t_{v})$ \;
        Get MSE on the validation set \;
         $S(\Delta) \leftarrow $MSE \;
      }
      $R(i) \leftarrow mean(S)$
   }
    $Best(omega) \leftarrow mean(R)$
 }
$\omega \leftarrow \arg\!\min_{\omega}(Best)$ \;
\end{algorithm}

\subsubsection{Variable kernel width}\label{k-nearest-section}

Rather than considering a fixed kernel width $\omega$,
in this section we allow variable width Gaussian kernels
of the form
\[
%\begin{equation}
%  \label{kernel2}
    K(c_{j},t_{i})=\exp{\frac{-|t_{i}-c_{j}|^{2}}{\omega_{j}^{2}}} \;;\;
    K(c_{j}+\Delta,t_{i})=\exp{\frac{-|t_{i}-(c_{j}+\Delta)|^{2}}
    {\omega_{j}^{2}}}.
%\end{equation}
\]
We determine each $\omega_{j}$ through a smoothing parameter $k \in
\{1,2,...,k_{max}\}$. Parameter $k$ is the
number of neighbouring observations $t_i$ on
both sides
of $c_j$ (boundary conditions need to be taken into account).
In particular, since we centre a kernel on each observation
time, i.e. $c_j=t_j$, we have the cumulative kernel width
\begin{equation}
  \label{kernel2}
%\omega_{j} = t_{j+d} - t_{j-d}.
\omega_{j}=\sum_{d=1}^{k}(t_{j} - t_{j-d})+(t_{j+d} - t_{j})=
\sum_{d=1}^{k}(t_{j+d} - t_{j-d}).
\end{equation}
The optimal value of $k$ can be estimated using a cross-validation procedure
analogous to that of section
\S \ref{CV-section}.

\begin{figure*}[htb!]
\centering
  \resizebox{\hsize}{!}{\includegraphics{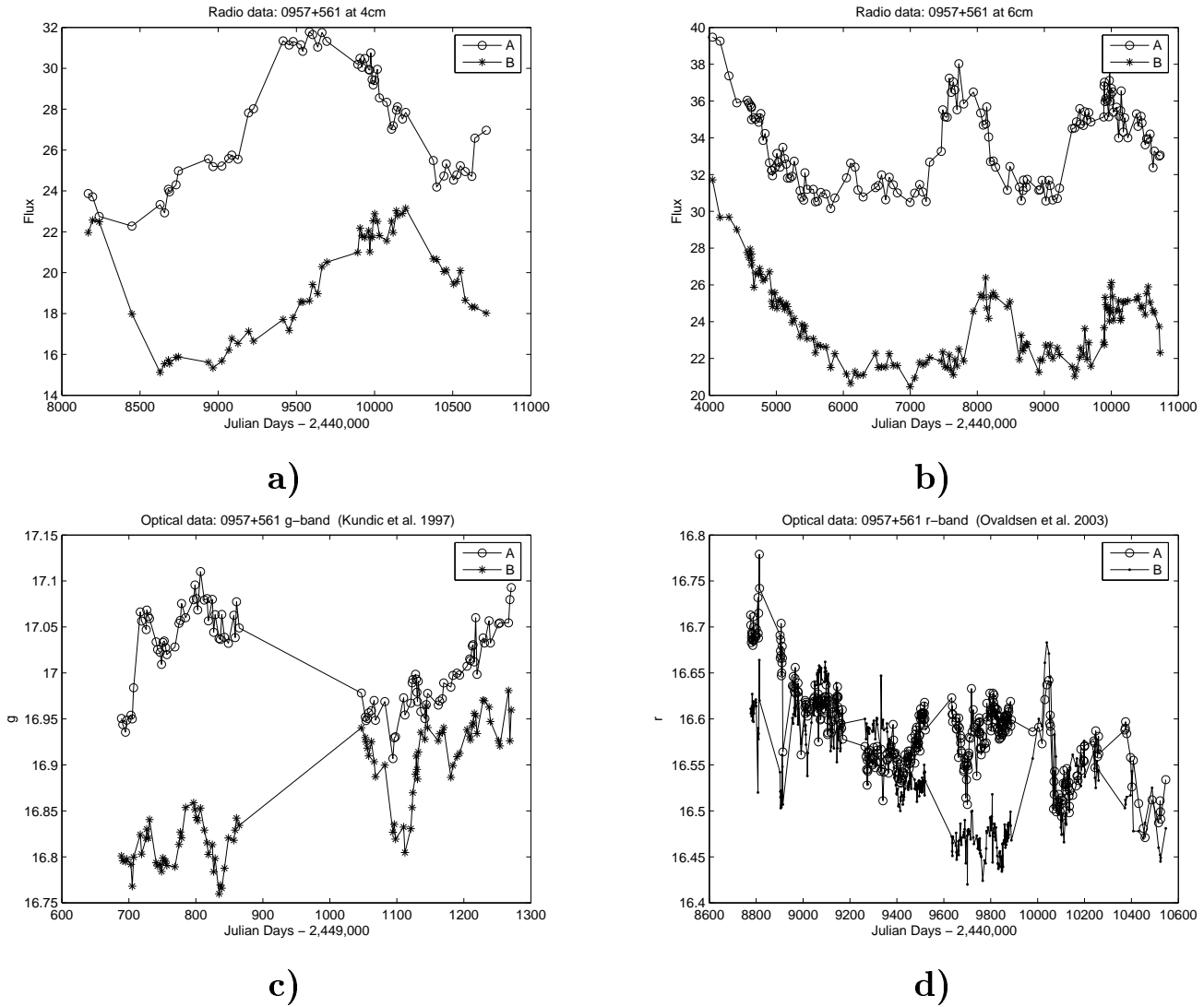}}
\caption{The variation of flux density with
time of the two gravitationally lensed images of
 Quasar \object{Q0957+561}. {\bf a)} Radio data at 4~cm, {\bf b)}
radio data at 6~cm \citep{Haarsma:1999:TRW}, {\bf c)} optical data at
g-band \citep{Kundic:1997:ARD}, and  {\bf d)}
optical data at r-band \citep{Ovaldsen:2003:NAP}.}
\label{0957-fig}
\end{figure*}

\section{Methods for estimating the Time delay}\label{methods-section}

Table~\ref{review_time_delay} contains a review, in chronological
order, of the more recent time delay estimates
of the quasar \object{Q0957+561} and the methods
employed. This gravitational lens is the most extensively
monitored so far, being the first one to be discovered. 
Fig. \ref{0957-fig} presents examples of the observed light curves
across various frequency bands, from radio to optical.
As is evident in
Table~\ref{review_time_delay},
whole range of time delay estimates
(with varying uncertainty bounds) for the gravitational lens
are available. The problem is that
{\em we do not know the actual
time delay}. One of the aims of this paper
is to study the reliability of several  time delay estimation methods
in a large set of controlled experiments on
artificially generated data with realistically modelled
observational noise and mechanisms of missing measurements.
We feel that only after learning lessons from such a study
does it make sense to come up with yet another batch of time delay
estimation claims.

\begin{table}
\caption{Review of time delay estimates between the two images of 
\object{Q0957+561} from 1997 to 2004. The methods are reviewed in \S\ref{methods-section}.}
\label{review_time_delay}
\centering
  \begin{tabular}{ l l l } \hline \hline
\textbf{Reference}&\textbf{Method(s)}&\textbf{Time delay}  \\ \hline
\citealt{Kundic:1997:ARD}  &- Linear              &   417$\pm$3 \\
                     & - Cross correlation  &   \\
                     &- PRH                 &   \\
                     &- Dispersion          &   \\
\citealt{Oscoz:1997:TDQ} &- Cross correlation   &   427$\pm$3 \\
                     &- Dispersion          &   \\
\citealt{Pijpers:1997:TDT} &- SOLA                &   425$\pm$17 \\
\citealt{Pelt:1998:MDT}&- Dispersion          &   416.3$\pm$1.7 \\
\citealt{Haarsma:1999:TRW}  &- PRH                 &   409$\pm$30  \\
                     &- Dispersion          &   \\
\citealt{Oscoz:2001:TDI} &- Linear              &   422.6$\pm$0.6\\
                     &- Cross correlation   &   \\
                     &- Dispersion          &   \\
\citealt{Burud:2001:ANA} &- $\chi ^{2}$ algorithm  &  423$\pm$9 \\
\citealt{Colley:2003:ATC} &- PRH                 &   417.09$\pm$0.07 \\
\citealt{Ovaldsen:2003:NAP} &- Dispersion          &   424.9$\pm$1.2 \\
                     &- $\chi ^{2}$ algorithm  &   \\ \hline
   \end{tabular}
\end{table}

In this section, we review the
principal time delay estimation methods that have been
used on gravitational lens data. The 
\textbf{Cross correlation} method \citep{Kundic:1997:ARD,Oscoz:1997:TDQ},
\textbf{PRH} method \citep{Press:1992:TTD} 
and \textbf{Dispersion} spectra,
\citep{Pelt:1996:TLC}, described in \S \ref{cross-correlation-section},
\S \ref{PRH-section} and \ref{dispersion-section}, respectively,
have been widely used in the literature.
We employ them in \S \ref{results-artificial-section}
as base-line models when reporting performance of our methods
(described in \S \ref{kernel-method-section}).

Of the methods mentioned in Table~\ref{review_time_delay}, the
\textbf{Linear} method uses chi-squared ($\chi ^{2}$) fitting
\citep[\S 14]{Press:1986:book}. Since the data are irregularly
sampled, linear interpolation in the observational gaps is performed
\citep{Kundic:1997:ARD}.

The method of Subtractive Optimally Localised Averages (\textbf{SOLA})
has been proposed as a method for solving inverse problems.  The
method was adopted by \citet{Pijpers:1997:TDT} who formulated time
delay estimation as an inverse problem.  It is worth nothing that SOLA
employs kernels, called averaging kernels. However, 
%our approach differs from SOLA 
{SOLA differs from our approach}
in several respects: {\bf (i)} SOLA does a symmetric
treatment of the two estimated fluxes
(flux A is fixed and flux B is varied to match A and vice versa),
{\bf (ii)} the reported time delay is the mean of the estimated
time delays in the two symmetric cases,
{\bf (iii)} a free parameter is used to adjust the
relative weighting of the errors in the variance-covariance matrix.
We also note that parameter estimation in SOLA is
problematic \citep{Larsen:1997:SOLA,Rabello:1999:SOLA} and this method
has been rarely used.

The \textbf{$\chi ^{2}$ algorithm}
\citep{Burud:2001:ANA,Ovaldsen:2003:NAP} is a $\chi ^{2}$-based method
similar in spirit to our model in that it also uses a notion of an
underlying model curve when fitting the two observed fluxes.
However, the underlying model is assumed to be regularly sampled.
It is regularised using a smoothing term
\citep[Eq. 3]{Burud:2001:ANA}.
Confidence intervals on the delay are estimated by performing
Monte Carlo simulations \citep{Burud:2001:ANA}. 

In general, when Monte Carlo simulations are not performed,
bootstrap techniques are used to calculate uncertainty in
time delay estimates.

\subsection{Cross correlation}\label{cross-correlation-section}

Basically, there are two versions of methods based on cross
correlation: the Discrete Correlation Function (DCF) and its variant, the
Locally Normalised Discrete Correlation Function (LNDCF). Both
calculate correlations directly on discrete pairs of light curves
\citep{Edelson:1988:DCF, Lehar:1992:TRT}. These methods avoid
interpolation in the observational gaps. Also, they are the simplest
and fastest time delay estimation methods.

First, time differences (lags), $\Delta t_{ij} = |t_j - t_i|$, 
between all pairs of observations 
are binned into discrete bins. Given a bin size $\Delta \tau$,
the bin centred at lag $\tau$ is the time interval 
$[\tau - \Delta \tau / 2,  \tau+\Delta\tau/2]$. 
$P(\tau)$ is the number of observational pairs in the bin 
centred at $\tau$.
The DCF at lag $\tau$ is given by
\begin{equation}
  \label{dcf}
   DCF(\tau) = \frac{1}{P(\tau)} \sum_{i,j} \frac{(x_A(t_i)- \bar{a})(x_B(t_j)- \bar{b}) }{\sqrt{(\sigma_a^2-\sigma_A^2(t_i))(\sigma_b^2-\sigma_B^2(t_j))} },
\end{equation}

\noindent 
where $\bar{a}$ and $\bar{b}$ are means of the observed data fluxes
$x_A(t_i)$ and $x_B(t_j)$, respectively; $\sigma_a^2$ and
$\sigma_b^2$ are their variances; $\sigma_A^2(t_i)$ and $\sigma_B^2(t_j)$
are the observational errors (\ref{error-model}).

Likewise,
\begin{equation}
  \label{lndcf}
   LNDCF(\tau) = \frac{1}{P(\tau)} \sum_{i,j} 
   \frac{(x_A(t_i)- \bar a(\tau))(x_B(t_j)-  \bar b(\tau)) }
   {\sqrt{(\sigma^2_a(\tau)-\sigma_A^2(t_i))
   (\sigma^2_b(\tau)-\sigma_B^2(t_j))} },
\end{equation}

\noindent 
where $\bar a(\tau)$, $\bar b(\tau)$, $\sigma^2_a(\tau)$ and
$\sigma^2_b(\tau)$ are the lag means and variances in the bin centred 
at $\tau$. 
The
time delay is found when $DCF(\tau)$ and $LNDCF(\tau)$
(\ref{dcf})-(\ref{lndcf}) are maximum, i.e. at the best correlation.

\subsection{The PRH method}\label{PRH-section}

This method is widely used for time delay estimation.
Its fundamentals are based on the theory of stochastic processes and
Wiener filtering \citep{Press:1992:TTD,Rybicki:1992:IRR}.
Given two light curves
$x_A$ and $x_B$ (\ref{data}), the 
PRH method combines them
into a single series $\vec{y}$ by assuming
a time delay $\Delta$ and a constant
ratio $M$ between $x_A$ and $x_B$.
Thus, for each of the two fluxes,
we end up having a new data set of $2n$ observations;
half is interpolated using the other flux.
The flux ratio $M$ is estimated as a difference
between weighted means
of the fluxes;
the weights are derived from the quoted observational errors.
The time delay, $\Delta$, is estimated by minimising
\begin{equation}
  \label{chi-square}
   \chi^{2}=\vec{y\/}^{\rm T} \left( \vec{A} - \frac{\vec{A} \vec{E} \vec{E\/}^{\rm T} \vec{A}}{\vec{E\/}^{\rm T} \vec{A} \vec{E}} \right) \vec{y},
\end{equation}
\noindent which is a measure of goodness of fit
on measurements from a Gaussian process \citep{Press:1992:TTD}.
Here, $\vec{y}$ is the combined flux\footnote{Note that \citet{Press:1992:TTD}
 refer to $\vec{y}$ as a component rather than combined components,
 image A and image B. The same occurs with the matrices $\vec{A}$,$\vec{B}$
 and $\vec{C}$ in Eq.~(\ref{A-matrix}) and (\ref{C-matrix}).},
 $\vec{E}$ is a column vector of ones, and
 \begin{equation} \label{A-matrix} \vec{A}=\vec{B}^{-1} \equiv \left\{
 C_{ab} + \langle \sigma_{a}^{2} \rangle \delta_{ab}
 \right\}^{-1}
\end{equation}
\noindent where
\begin{equation}
  \label{C-matrix}
       C_{ab}=\langle y(t_{a})y(t_{b}) \rangle  \equiv C(t_{a}-t_{b}) \equiv C(\tau)
  \end{equation}
\noindent is a covariance model estimated from the data\footnote{
Angle brackets denote the expectation operator.};
$t_{a}$, $t_{b}$, $a,b=1,...,2n$, are sample times of
the combined light curve.
\citet{Press:1992:TTD} suggest finding $C(\tau)$
through a first-order structure function $V(\tau)=\langle s^2 \rangle - C(\tau)$, where $s$ is the clean data from $\vec{y}$.
Then, the structure function $V(\tau)$
is computed from the data, single image, by determining lags
\begin{equation}
  \label{tau-ij}
      \tau_{ij} \equiv \vert t_{i}- t_{j} \vert
\end{equation}
and values
\begin{equation}
\label{v-ij}
v_{ij} \equiv ( x_{\{A,B\}}(t_{i})-x_{\{A,B\}}(t_{j}))^{2}-\sigma_{\{A,B\}}^{2}(t_{i})-\sigma_{\{A,B\}}^{2}(t_{j}) .
\end{equation}
\noindent where $\{A,B\}$ denotes that it comes from either image A or image B (\ref{data})-(\ref{error-model}).

All pairs  ($\tau_{ij}$, $v_{ij}$) are sorted with respect $\tau_{ij}$ and
binned into 100 bins \citep[pg. 407]{Press:1992:TTD}.
The values of $\tau_{ij}$ and $v_{ij}$ in each bin are averaged and
finally a power-law model is built to fit the binned list,
\begin{equation}
  \label{power-law-model}
  V(\tau)=c_{1}\tau^{c_{2}}.
\end{equation}
Note that this model is linear in log scale,
\begin{equation}
  \label{log-power-law-model}
  V(\ln(\tau))=\ln(c_{1})+{c_{2}}\ln(\tau).
\end{equation}
Parameters $c_{1}$ and $c_{2}$ of the structure function can be
determined using a simple line fitting algorithm\footnote{
We have noticed that in some cases a negative slope $c_2$ is found. Also be aware that
a negative $c_{1}$, $y$-intercept, in Eq. (\ref{log-power-law-model}), and
$\tau = 0$ leads to numerical overflow.
In such cases we apply a shift up in Eq. (\ref{v-ij}), and we set $\tau$ to a very small positive number.}.
So, $V(\tau)$ is estimated on a single flow and one would naturally expect
that estimates of $V(\tau)$ on flux $A$ would be similar to those
on flux $B$. However, this is often not the case.
\citet{Press:1992:TTD}
claim that it does not matter which image is chosen for the
$V(\tau)$ estimation as the time delay calculations are sufficiently
robust to variations in the $V(\tau)$ estimates.
Our experience suggests that this may be an overoptimistic
expectation.
Moreover, the matrix $\vec{B}$ (\ref{A-matrix})
is often ill conditioned and we regularise the inversion operation through
SVD.

\subsection{Dispersion spectra}\label{dispersion-section}

Dispersion is a weighted sum of
squared differences between $x_{A}(t_{i})$ and $x_{B}(t_{i})$
\citep{Pelt:1996:TLC, Pelt:1998:MDT,
Pelt:1998:EMT, Pelt:2002:BCT}. 
The method is similar to those based on DCF (see \S
\ref{cross-correlation-section}).
However, it models the time series
of two light curves in a different way by combing them
(given a time delay $\Delta$ and ratio $M$) into a single flux flow,
$\vec{y}$, as in the PRH method (\S \ref{PRH-section}).
We worked with two versions of
this method \citep[see][]{Pelt:1998:MDT}:

\begin{equation}
  \label{D1-equation}
   D_{1}^{2}(\Delta) = \!\min_{M} \frac{\sum_{a=1}^{2n-1}w_{a} \left( y(t_{a+1})-y(t_{a}) \right) ^{2}}{2\sum_{a=1}^{2n-1}w_{a}}
\end{equation} 

\noindent and

\begin{equation}
  \label{D4-equation}
   D_{4,2}^{2}(\Delta) = \!\min_{M} \frac{\sum_{a=1}^{2n-1}\sum_{c=a+1}^{2n}S_{a,c}^{(2)}W_{a,c}G_{a,c} \left( y(t_{a})-y(t_{c}) \right) ^{2}}{2\sum_{a=1}^{2n-1}\sum_{c=a+1}^{2n}S_{a,c}^{(2)}W_{a,c}G_{a,c}},
\end{equation}
where

\begin{equation}
  \label{W-equation}
   w_{a}=\frac{1}{\sigma^2(t_{a+1})+\sigma^2(t_{a})}, \\  W_{a,c}=\frac{1}{\sigma^2(t_{a})+\sigma^2(t_{c})} 
\end{equation}

\noindent are the statistical 
weights taking in account the measurement errors 
(\ref{error-model}). 
$G_{a,c}=1$ only when $y(t_{a})$ and $y(t_{c})$ are 
from different images, and $G_{a,c}=0$ otherwise.

\begin{equation}
  \label{S-equation}
   S_{a,c}^{(2)}= \left\{ \begin{array}{ll}
                                1-\frac{\vert t_{a} - t_{c} \vert}{\delta}, & \hbox{if}\ \ \vert t_{a} - t_{c} \vert \leq \delta\\
                                0, & \hbox{otherwise}.
                          \end{array}
                  \right.
\end{equation}

\noindent The estimated time delay, 
$\Delta$, is found by minimising
$D^{2}$ over a range of time delay trials.

Compared with $D_{1}^{2}$, the  
$D_{4,2}^{2}$ method has an additional parameter, 
{\it decorrelation length} $\delta$, that
signifies the 
maximum distance between observations we are willing to consider 
when calculating the correlations \citep{Pelt:1996:TLC}.

\section{Constructing Artificial data sets}\label{artificial-section}

We use artificial data sets to perform a set of {\em controlled} \
large-scale experiments in
order to measure the accuracy of time delay estimation techniques on
gravitational lens systems. We generate simulated data sets with
different levels of noise and
varying sizes/locations of
observational gaps.

The basic signal is constructed
by superimposing $G=20$
Gaussian functions with centres and widths generated randomly. 
The width is allowed to vary from zero up to
a quarter of the duration of the entire monitoring campaign.
Then, two artificial fluxes are created by scaling and shifting
the basic signal in the flux density and time domains, respectively.
The amplitude and flux densities are similar to radio data, 4~cm
\citep{Haarsma:1999:TRW}. The flux ratio was set to
$M=1/1.44$ and the temporal shift was equal to $\Delta=500$ days.
The time
goes from 0 to $T\cdot\Delta$ days with $s_{1}$ samples per $\Delta$ days ($T=10$ and
$s_{1}=5$), i.e. if the samples were regularly sampled, we would
have a separation of $z=\Delta/s_{1}$ days 
between samples. To irregularly
sample, we disturb the regular observation times
with a random variable uniformly distributed in
$[-P\cdot z,
+P \cdot z]$, $P=0.49$.
Moreover, we simulated continuous gaps in observations
by imposing $g=5$ blocks of missing data.
The blocks are located randomly with
at least one sample between them.
We worked with block lengths $s_{2}=1,2,...,5$ (see Table \ref{data-sets-table}).

\begin{figure}[ht!]
\centering
\resizebox{\hsize}{!}{\includegraphics{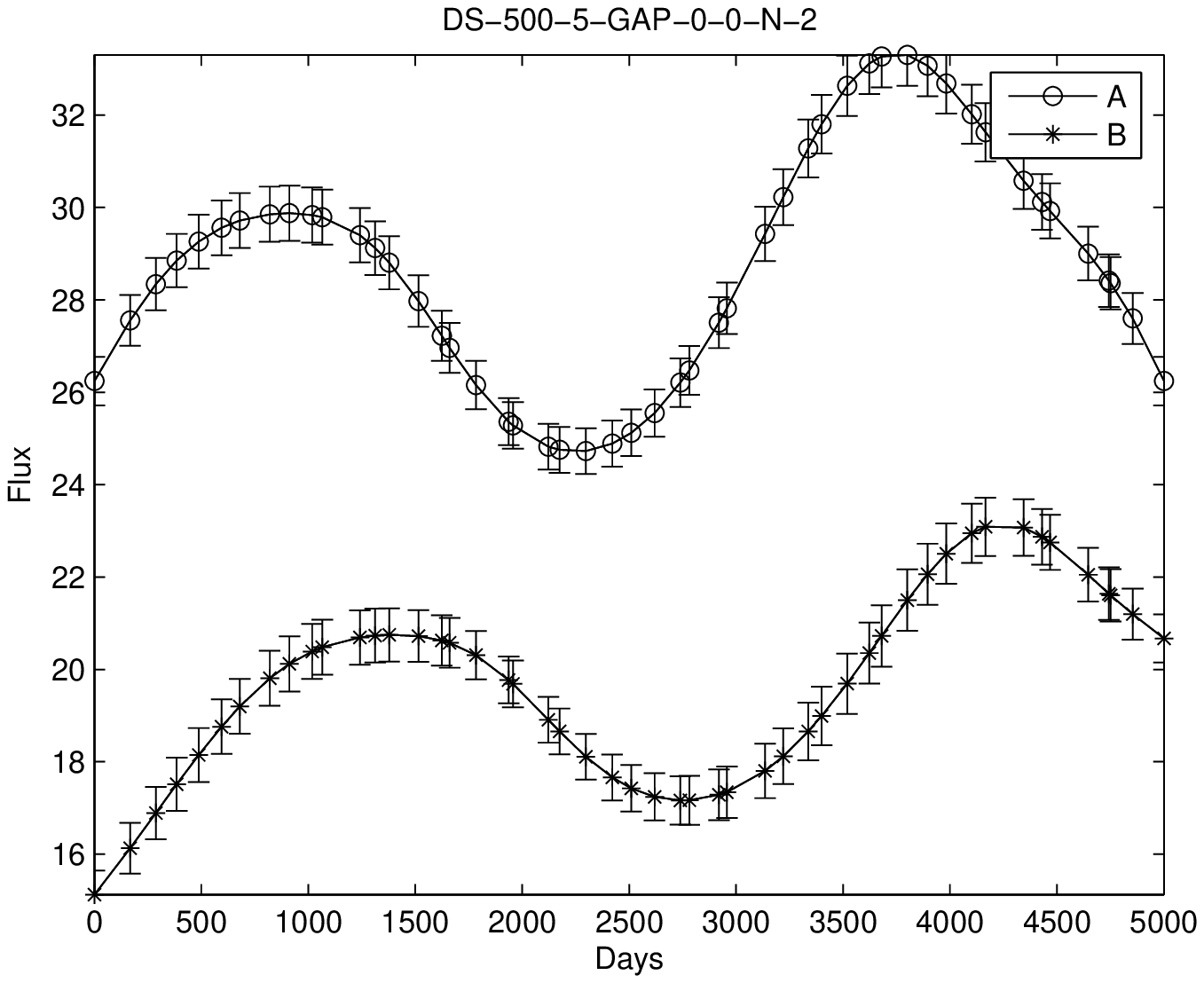}} \\
\resizebox{\hsize}{!}{\includegraphics{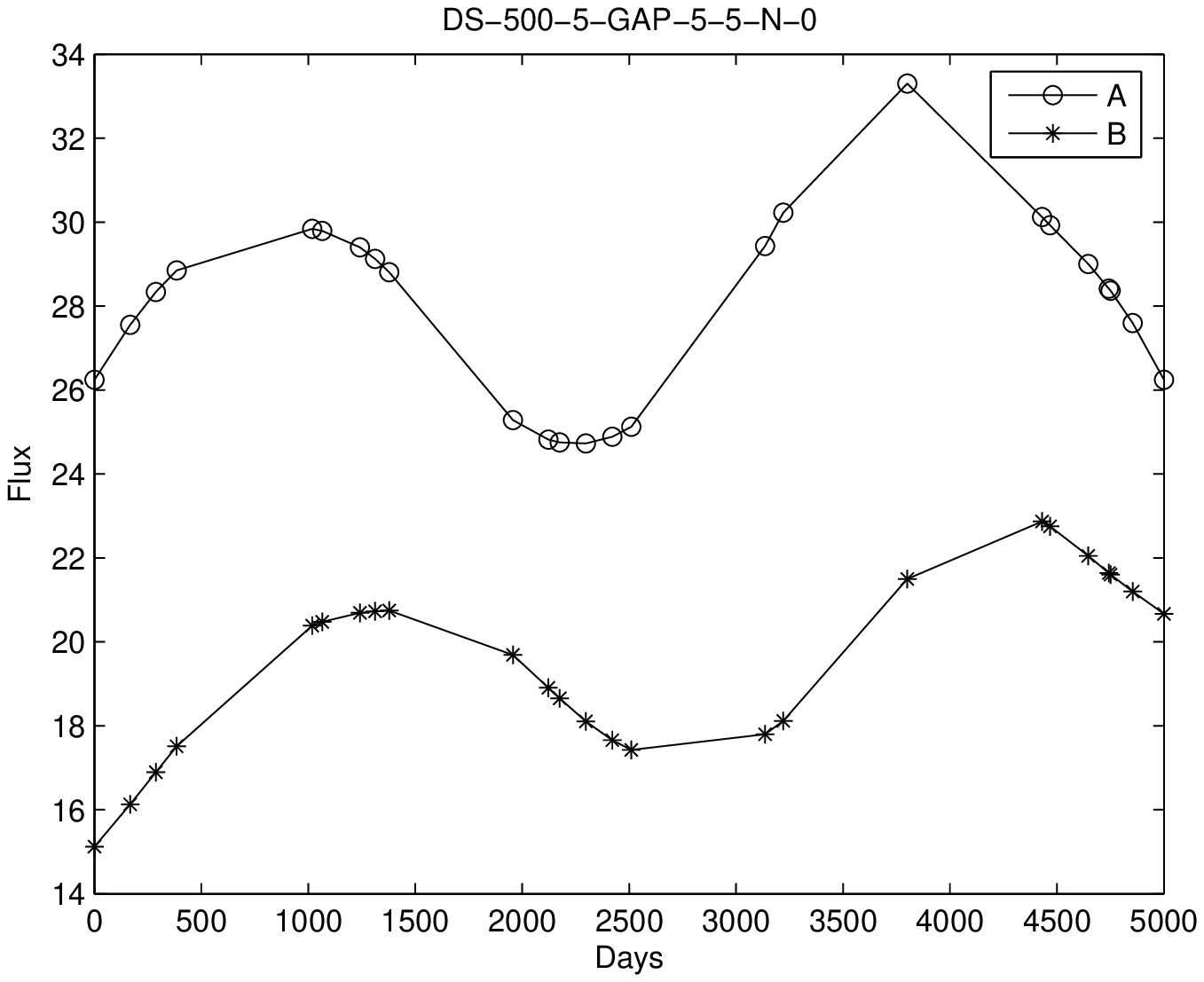}}
\caption{Artificial flux data generated to
simulate a couple of scaled and shifted fluxes coming from
a quasar through a gravitational lens.
The top plot shows the underlying function without
observational gaps. Also shown are the error bars of 2\% of the flux
value. Below are the same noise-free fluxes
with imposed observational gaps of length 5.}
\label{artificial-fig}
\end{figure}

Three levels of noise were used to contaminate the flux signal:
1\%, 2\% and 3\% of the flux;
these represent our measurement errors
$\sigma_{A}(t_{i})$ and $\sigma_{B}(t_{i})$,
which are standard deviations of the flux distribution
at each observation time 
(see  
Eqs. (\ref{data}) and (\ref{error-model})).
Fig. \ref{artificial-fig} shows an example of
a couple of scaled and shifted artificial fluxes\footnote{More plots
can be found at \\ http://www.cs.bham.ac.uk/$\sim$jcc/artificial/ }.

We used $20$ different underlying functions (basic signals).  For each underlying function,
we generated $100$ realisations for each noise level by adding
a Gaussian noise to the underlying function as in equations
(\ref{data}) and (\ref{error-model}).
For each such data set,
we performed $10$ realisations of missing observational blocks.
Overall, we employed $307\,020$ different data sets,
$15\,351$ data sets per underlying function
(see Table \ref{data-sets-table}).
\begin{table}
\caption{Artificial data sets under analysis}
\label{data-sets-table}
\begin{center}
  \begin{tabular}{ c | l l l l l l } \hline \hline
                &\multicolumn{5}{c}{Gap size $s_2$} & \\ \hline
  \textbf{Noise}&\textbf{0}&\textbf{1}&\textbf{2}&\textbf{3}&\textbf{4}&\textbf{5} \\  \hline
      0\%  & 1   & 10   & 10   & 10   & 10   & 10    \\
      1\%  & 100 & 1\,000 & 1\,000 & 1\,000 & 1\,000 & 1\,000  \\
      2\%  & 100 & 1\,000 & 1\,000 & 1\,000 & 1\,000 & 1\,000  \\
      3\%  & 100 & 1\,000 & 1\,000 & 1\,000 & 1\,000 & 1\,000  \\ \hline
 Sub-Total & 301 & 3\,010 & 3\,010 & 3\,010 & 3\,010 & 3\,010 \\ \hline
   \multicolumn{7}{l}{Total = 15\,351 data sets per underlying function.} \\
   \multicolumn{7}{l}{20 underlying functions yield 307\,020 data sets.} \\
 \end{tabular}
\end{center}
\end{table}

\section{Experiments with artificial data}\label{results-artificial-section}
\begin{figure}[h]
\centering
  \resizebox{\hsize}{!}{\includegraphics{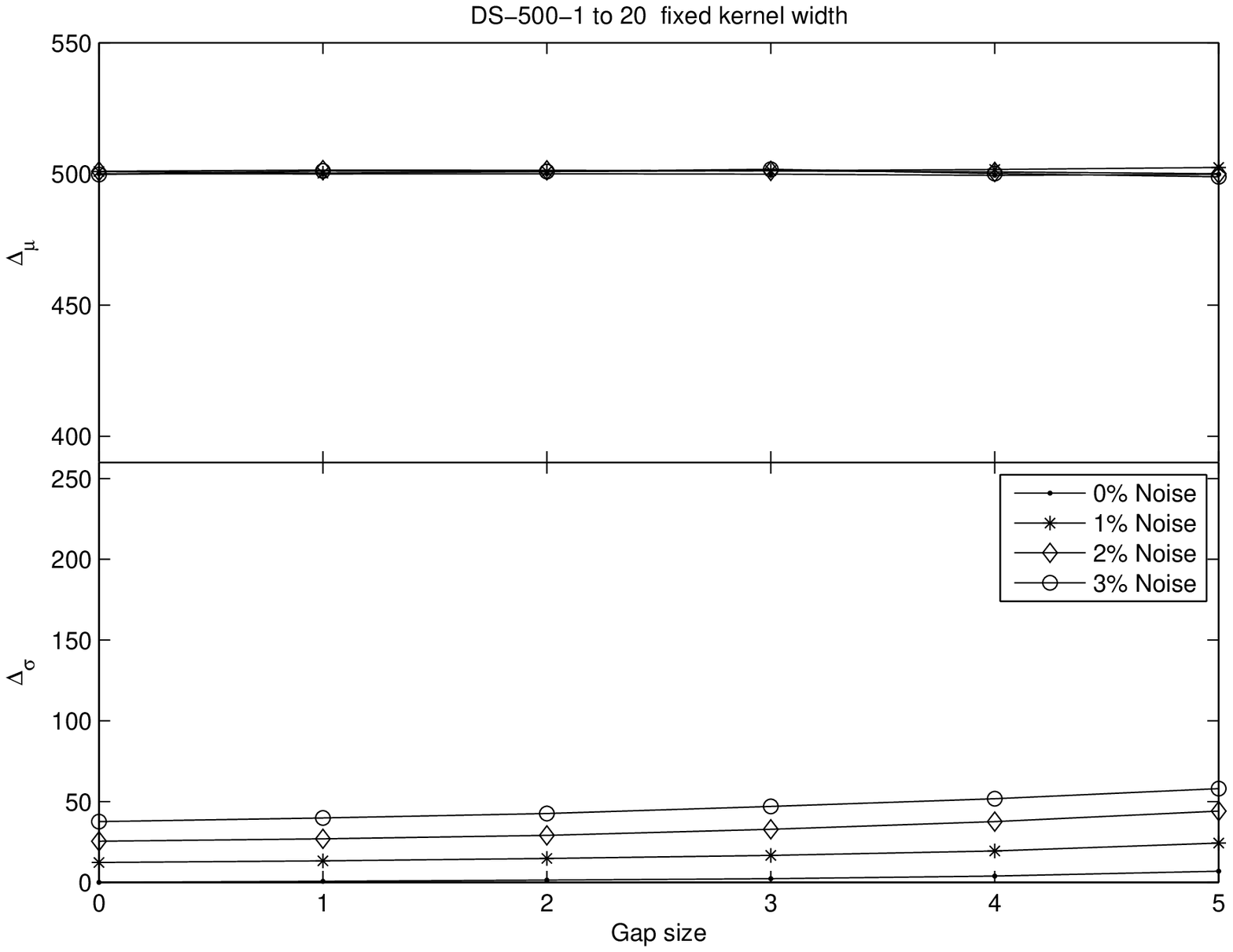}}
\caption{Results of the application
of our Kernel method with fixed width (in \S \ref{CV-section}) on all
artificial data sets (see \S \ref{artificial-section}). Details in \S
\ref{results-artificial-section}.}
\label{cv-fig}
\end{figure}

\begin{figure}[h]
\centering
  \resizebox{\hsize}{!}{\includegraphics{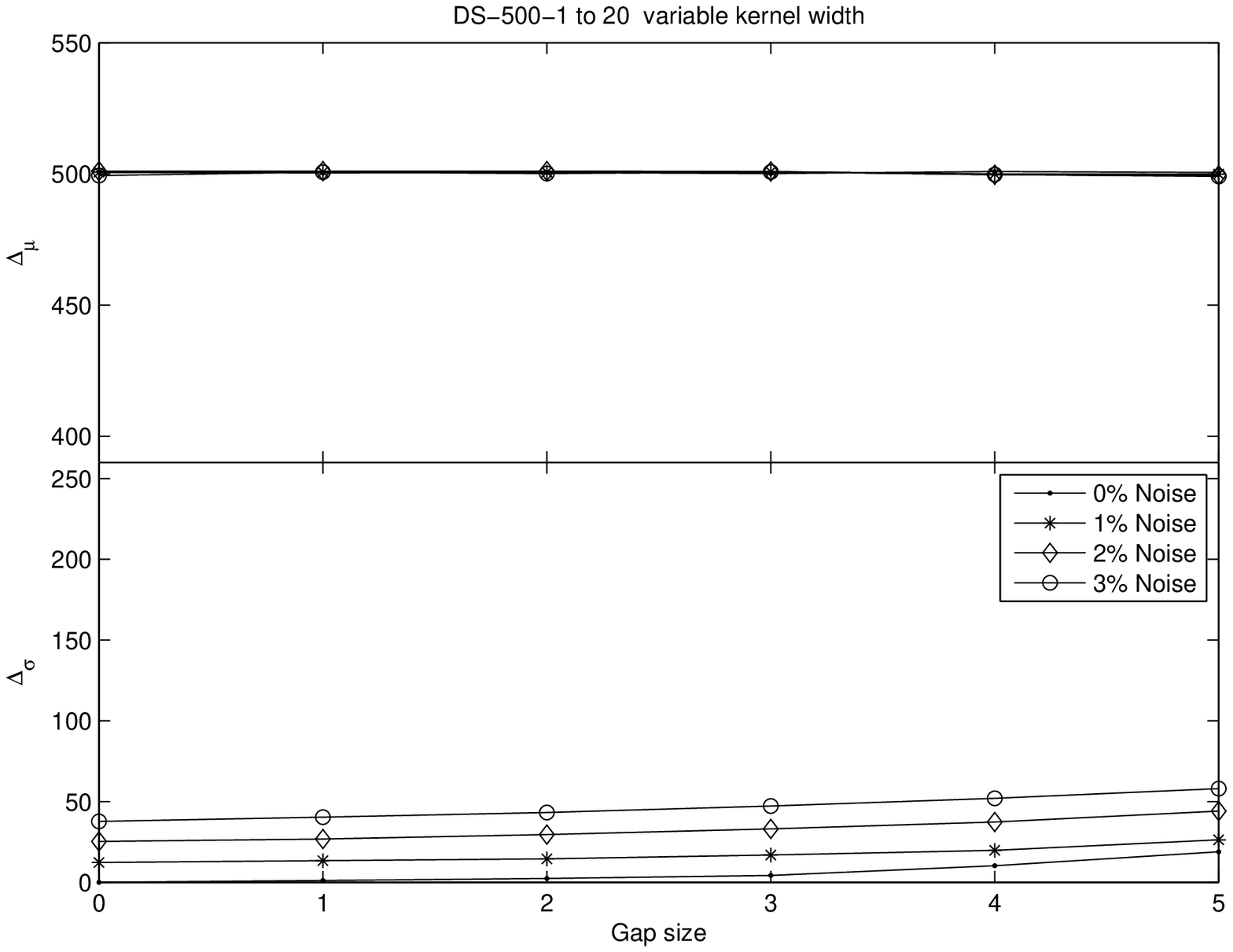}}
\caption{Results of the application
of our Kernel method with variable width (in \S
\ref{k-nearest-section}) on all artificial data sets (see \S
\ref{artificial-section}). Details are in \S
\ref{results-artificial-section}. }
\label{k-nearest-fig}
\end{figure}

\begin{figure}[h]
\centering
  \resizebox{\hsize}{!}{\includegraphics{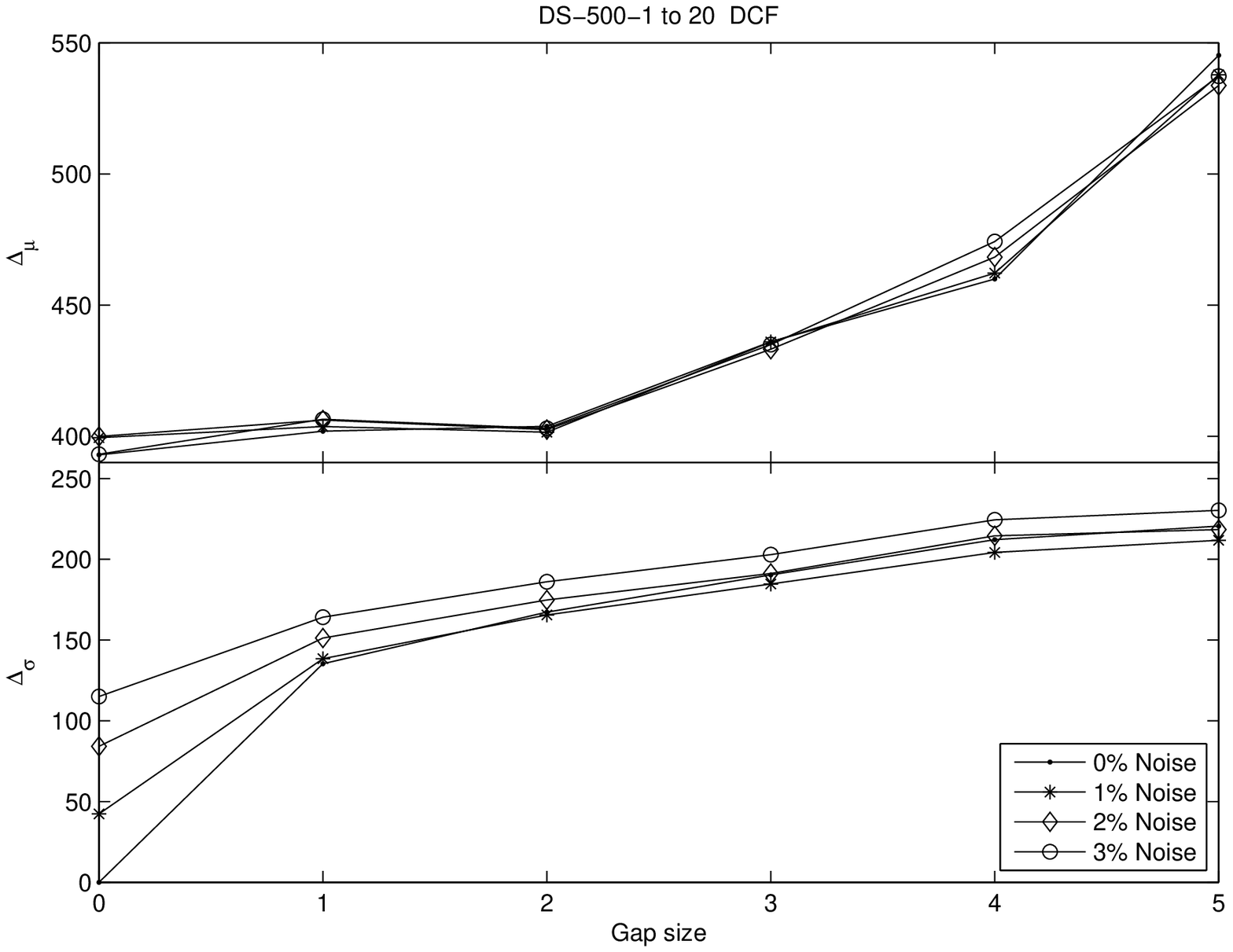}}
\caption{Results of the application of the DCF method (in \S \ref{cross-correlation-section}) on all artificial data sets (see \S \ref{artificial-section}). Details are in \S \ref{results-artificial-section}.}
\label{dcf-fig}
\end{figure}

\begin{figure}[h]
\centering
  \resizebox{\hsize}{!}{\includegraphics{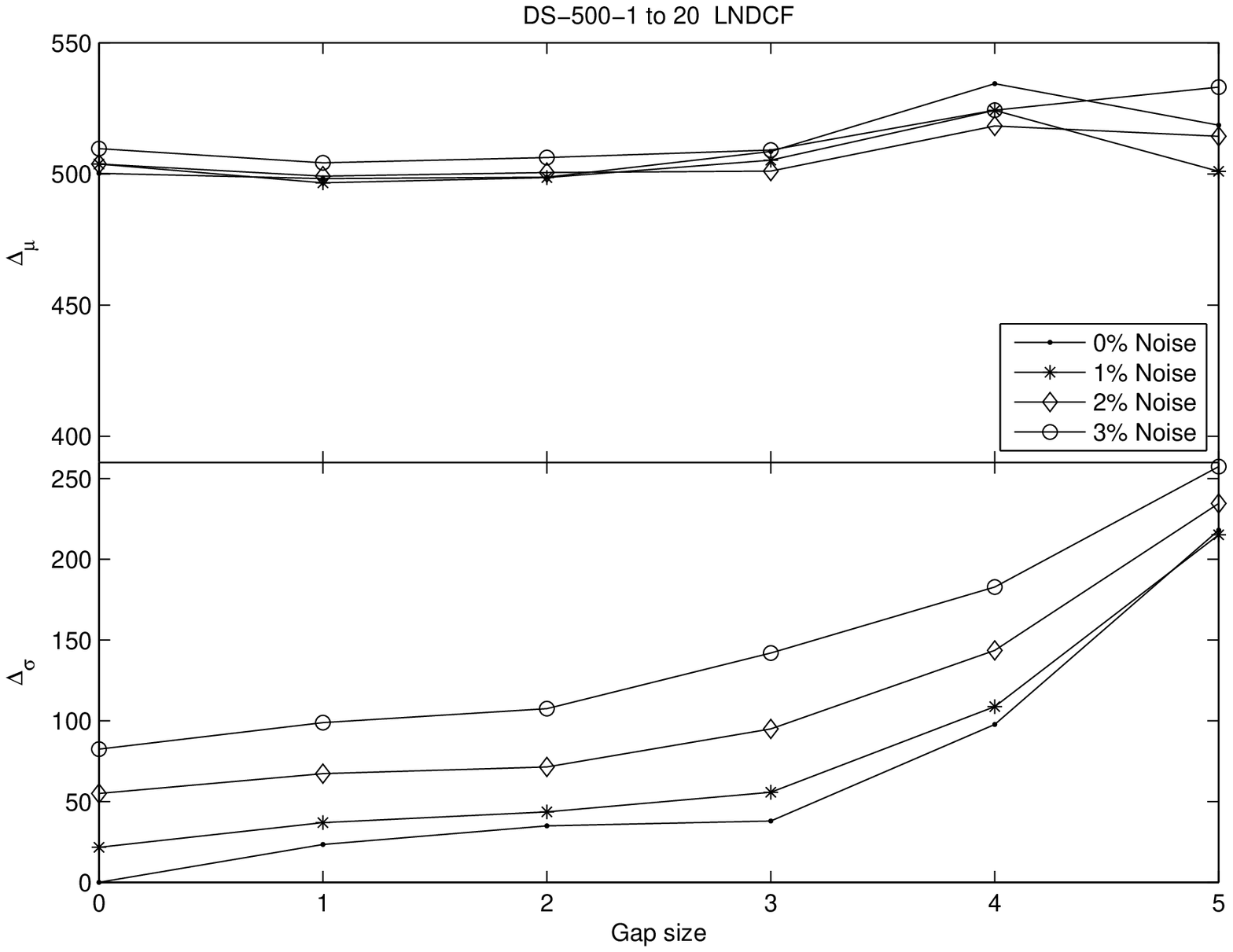}}
\caption{Results of the application of the LNDCF method 
(in \S \ref{cross-correlation-section}) on all artificial data sets
(see \S \ref{artificial-section}). Details are in \S
\ref{results-artificial-section}.}
\label{lndcf-fig}
\end{figure}

\begin{figure}[h]
\centering
  \resizebox{\hsize}{!}{\includegraphics{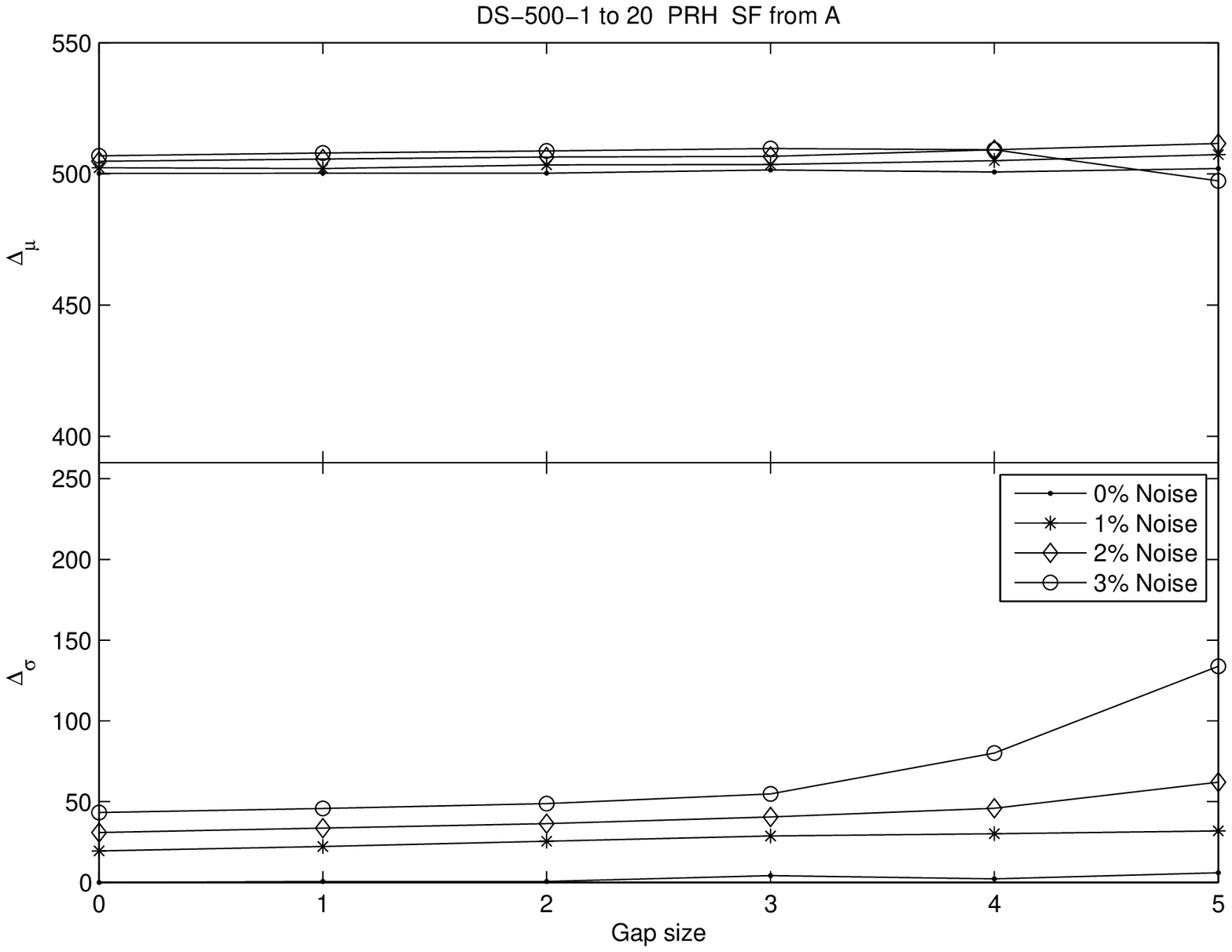}}
\caption{Results of PRH method with structure function from image A (in \S \ref{PRH-section}) on all artificial data sets (see \S \ref{artificial-section}) except those cases where negative slope occur (355 cases). Details in \S \ref{results-artificial-section}.}
\label{prh-a-fig}
\end{figure}

\begin{figure}[h]
\centering
  \resizebox{\hsize}{!}{\includegraphics{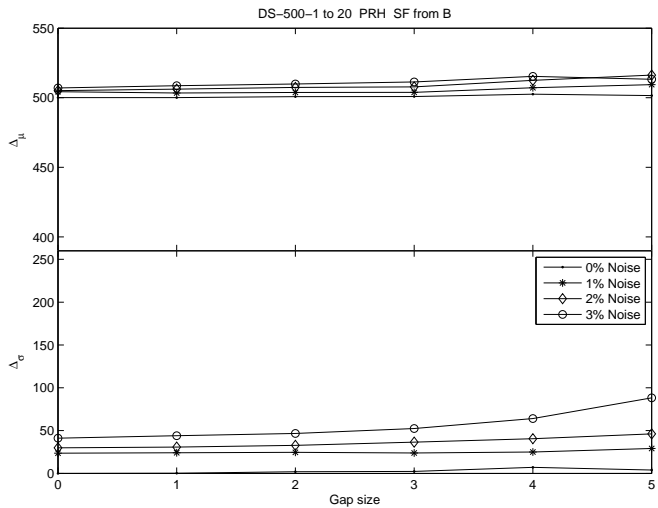}}
\caption{Results of the application of the 
PRH method with structure function from image B (in \S \ref{PRH-section}) on all artificial data sets (see \S \ref{artificial-section}) except those cases where negative slope occur (105 cases). Details are in \S \ref{results-artificial-section}.}
\label{prh-b-fig}
\end{figure}

\begin{figure}[h]
\centering
  \resizebox{\hsize}{!}{\includegraphics{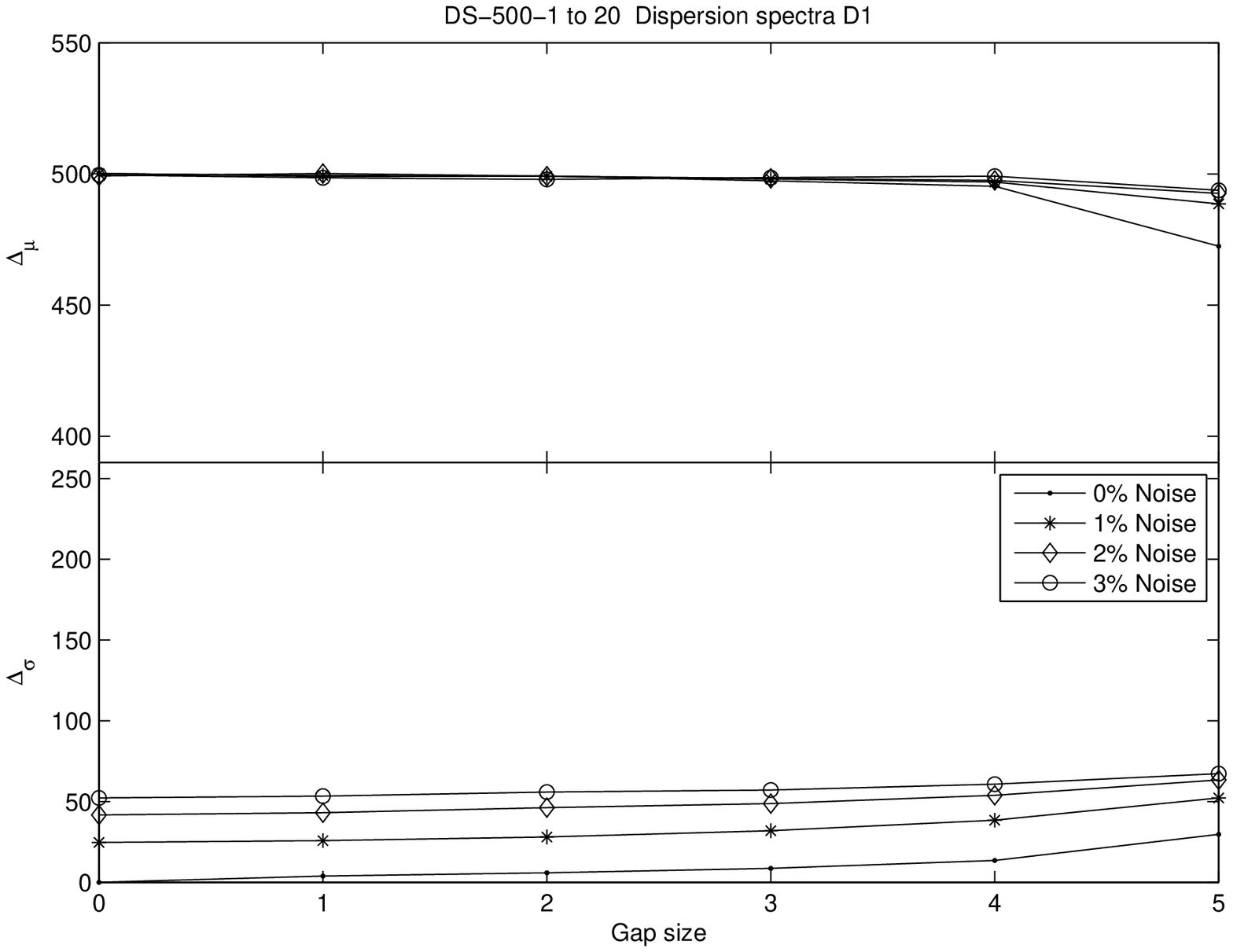}}
\caption{Results of the application of the Dispersion spectra 
method $D_{1}^{2}$ (in \S \ref{dispersion-section}) on all artificial data sets (see \S \ref{artificial-section}). Details are in \S \ref{results-artificial-section}.}
\label{d1-fig}
\end{figure}

\begin{figure}[h]
\centering
  \resizebox{\hsize}{!}{\includegraphics{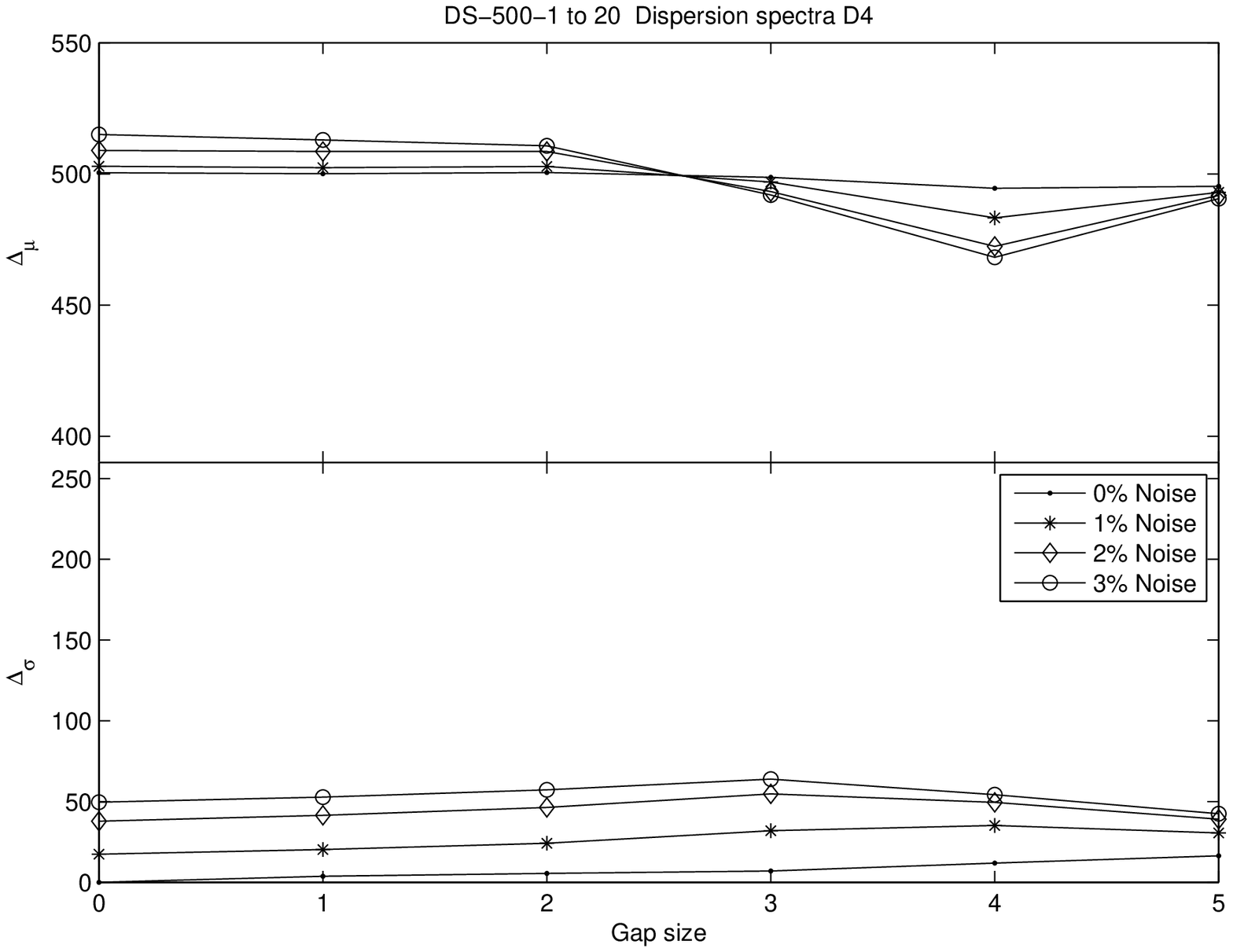}}
\caption{Results of the application of
the Dispersion spectra method $D_{4}^{2}$ (in \S
\ref{dispersion-section}) on all artificial data sets (see \S
\ref{artificial-section}). Details are in \S
\ref{results-artificial-section}.}
\label{d4-fig}
\end{figure}

In this section, we test our methods of \S \ref{CV-section} and
\S \ref{k-nearest-section} on artificial data sets
described in \S \ref{artificial-section} and
compare them to the existing methods described in
\S \ref{cross-correlation-section}, \S \ref{PRH-section} and
\S \ref{dispersion-section}.
Figures \ref{cv-fig}--\ref{d4-fig} show
two kinds of curves. The
top panel shows curves representing the mean estimated time delay
$\Delta_{\mu}$
versus the gap size for different noise levels, while
lower curves represent means of standard deviations $\Delta_{\sigma}$ of the estimated
time delay per underlying function.
The quantities of data sets involved in this analysis are 
shown in
Table~\ref{data-sets-table}.

In all experiments reported in this section, the following parameter
settings were used: $M=1/1.44$, $\Delta_{min}=400$ and
$\Delta_{max}=600$ with increments of $1$ day.  We used a threshold of
$0.001$ (found empirically) to regularise inversion in
Eq. \ref{inverse} through SVD, discarding singular values less than the
threshold \citep[\S 2]{Press:1986:book}.

Results for the fixed kernel width technique (Algorithm \ref{cross-validation})
are shown in Fig. \ref{cv-fig}. Here,
${\it LowerBound}\!=\! 900$ and ${\it UpperBound}\!=\! 1200$ with increments of $10$.
Fig. \ref{k-nearest-fig} shows results of the variable kernel width technique.
We fixed the number of neighbours to $k=3$, which was estimated 
through cross-validation (see Algorithm \ref{cross-validation}) 
with ${\it LowerBound}\!=\! 1$ and 
${\it UpperBound}\!=\! 15$ with increments of $1$.

Figures~\ref{dcf-fig} and \ref{lndcf-fig} contain results for the DCF
and LNDCF methods respectively. For both methods, bin size of $100$
days (which is close to the lag average) was used and a search was
performed for the maximum correlation on bins in the range of $0$ to
$2\Delta$ days ($\Delta=500$ for artificial data).  We tested DCF
and LNDCF on regularly sampled data sets as in \S
\ref{artificial-section} (each bin contains those pairs with the
same lag). We found that LNDCF never fails, but DCF fails for some
shapes of the underlying function (e.g. flat shapes). So, we recommend the use of LNDCF, in
preference to DCF.

Figure~\ref{prh-a-fig} displays the results of the PRH method using image
A to estimate the structure function (\ref{power-law-model}), while
Fig.~\ref{prh-b-fig} shows results obtained using structure function
estimated on image B.  When estimating the structure function for each
data set in Table \ref{data-sets-table}, we use bins in the range
$100-700$ days \citep{Haarsma:1999:TRW}.  Linear regression was used
to estimate parameters $c_{1}$ and $c_{2}$ in
(\ref{log-power-law-model}). As pointed out in \S \ref{PRH-section},
some artificial data sets yield a negative slope due to gaps, high
noise and flat features on some underlying functions (when $s_2 \geq
3$ and noise $\geq 2$\%). Therefore, in such cases, we omit them to
get more reliable results. Also, we regularise as above to invert
$\vec{B}$ (\ref{A-matrix}), because zero noise and duplicate times may
occur in $\vec{y}$ (\ref{chi-square}) leading to
singularity. Consequently, we do not use the fast methods to get
$\vec{A}$ in (\ref{A-matrix}) \citep[see][]{Rybicki:1995:CFM}.

The results of the Dispersion spectra method are in Figs. \ref{d1-fig} and
\ref{d4-fig} for $D_{1}^{2}$ and $D_{4,2}^{2}$ respectively.  We set
$\delta=100$ as decorrelation length
\footnote{This value of decorrelation length gives the best
resolution on artificial data. \citet{Pelt:1996:TLC} and
\citet{Haarsma:1999:TRW} used $\delta=60$ for radio data.} 
for $D_{4,2}^{2}$.

We point out that the results in Figs. \ref{cv-fig} to \ref{d4-fig}
were obtained on the same collection of artificial data sets
and are plotted with the same scale on the {\it y}-axis.
Compared to the existing methods (DCF, LNDCF, PRH, Dispersion spectra),
our methods are more accurate and robust
with respect to the increasing gap size and noise level.
In general, for all methods,
there is an obvious tendency of increased uncertainty as the gap size increases. Increasing noise levels in the data result in increased uncertainty
of the time delay estimates.

\begin{figure*}[p!]
\centering
\includegraphics[width=17cm]{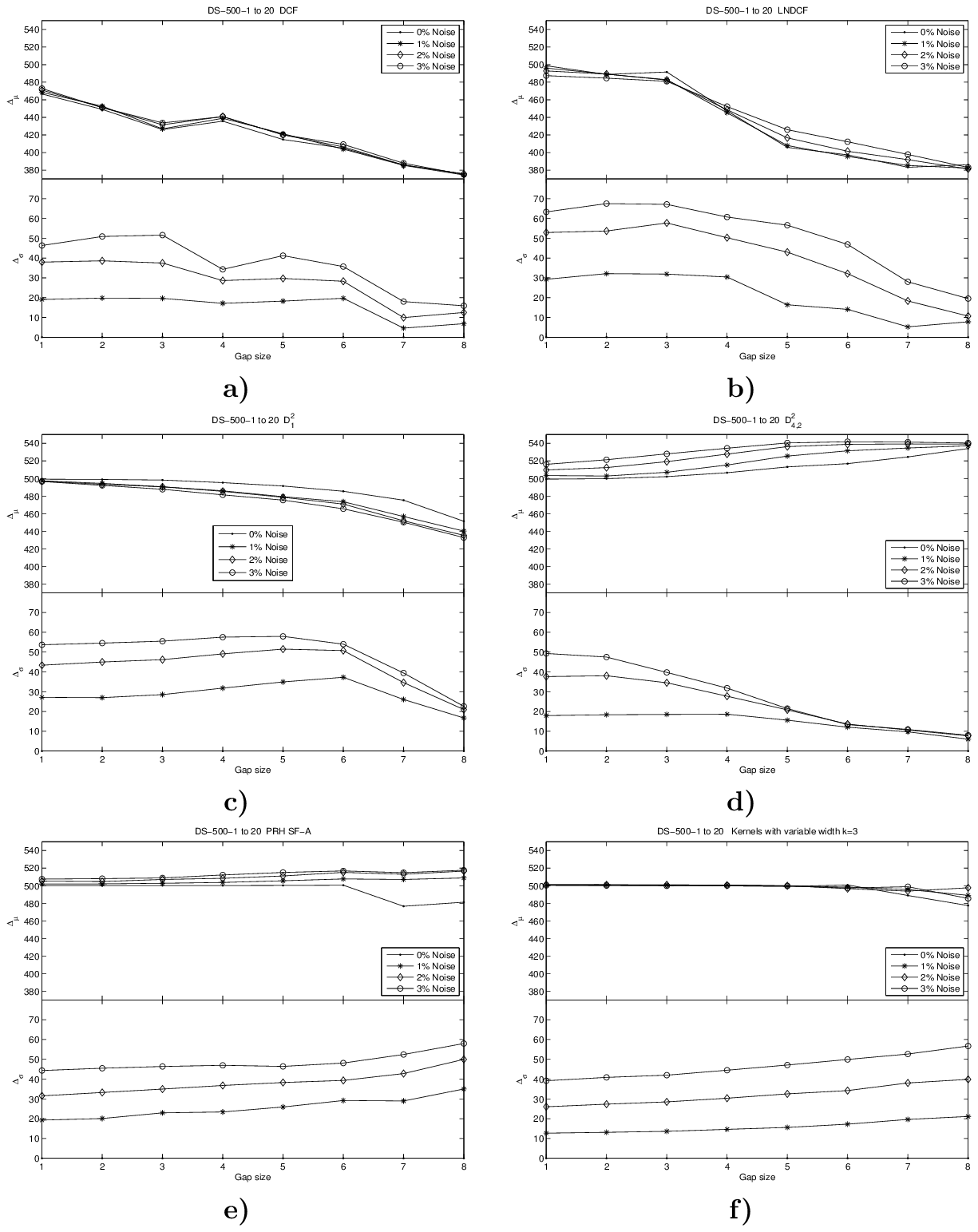}
\caption{Results on artificial data sets with periodic gaps, see \S \ref{artificial-section} and \S \ref{results-artificial-section}. All plots have the same {\it y}-axis scale, and the results are on the same data sets but with different method: {\bf a)} DCF with bin size $\Delta\tau=100$ (see \S \ref{cross-correlation-section}), {\bf b)} LNDCF with bin size $\Delta\tau=100$ (see \S \ref{cross-correlation-section}), {\bf c)} Dispersion spectra $D_1^2$ (see \S \ref{dispersion-section}). {\bf d)} Dispersion spectra $D_{4,2}^2$ with $\delta=100$ (see \S \ref{dispersion-section}), {\bf e)} PRH method with structure function from image A (see \S \ref{PRH-section}), and {\bf f)} Kernels with variable width $k=3$ (see \S \ref{k-nearest-section}). The results are similar to those in Figs. \ref{k-nearest-fig} to \ref{d4-fig}, respectively. See \S \ref{results-artificial-section} for more details.}
\label{periodic-fig}
\end{figure*}

We also tested our method on (annual) periodic gaps\footnote{We are thankful to the anonymous
reviewer for making this suggestion.} ranging from 1 to 8 months, corresponding to observing seasons from 4 to 11 months.
The results are in Fig. \ref{periodic-fig} f). The performance of DCF, LNDCF, Dispersion spectra and PRH method (structure function from image A only) on periodic gaps is also depicted in Fig. \ref{periodic-fig}. The parameters of methods were estimated as above, except for DCF and LNDCF where now we look for a maximum correlation between 400 and 600 days (showing less variance $\Delta_{\sigma}$); sometimes the time delay is below 400 days because we adopt a lower bin if there is no bin of 400 days. We see again that our method outperforms others.

\section{Artificial data sets through PRH method}\label{artificial-prh-section}

A possible criticism of our testing framework in \S \ref{artificial-section} may be that
we construct artificial underlying functions (fluxes)
as linear superpositions of Gaussian functions, while our proposed model
is a linear superposition of Gaussian kernels (see
(\ref{data})--(\ref{kernel})).  
However,  widths of the Gaussian functions used to construct 
the underlying functions
are much larger than widths of Gaussian kernels in our model
formulation, and so this criticism is less relevant.
Still, in order to properly address this issue, 
we let the PRH method "play at its own game" by constructing
a set of underlying functions using PRH method\footnote{We are thankful to the anonymous
reviewer for making this suggestion.} with a
specified structure function (SF).
We refer to such data sets as artificial PRH data. 
These are Monte Carlo time series, generated exactly as
described in \citet[][\S 5.2]{Press:1992:TTD}, with a fixed SF
given by $c_1=1/5.36 \times 10^5$ and $c_2=0.246$
\citep{Vanden_Berk:2004:TEP,Pindor:2005:DGL}. We use a monitoring
campaign length of 8 months with different sampling rates: ({\bf i}) Low,
every seven days. ({\bf ii}) High, every three days. ({\bf iii}) Irregular, every
two days with periodic gaps of fifteen days
\citep{Eigenbrod:2005:TCM}. ({\bf iv}) Irregular, as in \S
\ref{artificial-section} with lags of 3 days. 

We randomly choose seven
time delays in the range of 30--100 days. For each combination of
time delay and sampling rate, we generate 100 Monte Carlo data
sets. To simulate observational errors, we use fixed variances of $1
\times 10^{-7}$ and $1 \times 10^{-9}$ in order to get distinguishable
shapes by eye, i.e. low noise. Then, we analyse all the 
artificial PRH data sets  with our
methods in \S \ref{kernel-method-section} and the PRH method as
described in \S \ref{PRH-section}. 

If we keep the SF (i.e. $c_1$ and $c_2$) fixed to its true value
(as assigned to the simulations), then the performance of the PRH method
is outstanding, with almost zero bias and zero variance
for all data sets as a whole.  However, the performance is not always
as good when applied to individual data sets (one realisation). 
Further, if we assume that we do not know the true
SF used to generate the data, and we estimate it
through the data itself (see \S \ref{PRH-section}), then our methods perform better than the PRH method, as shown 
in the case of the artificial data in \S \ref{artificial-section} (see Table \ref{PRH-data-results}).
In fact, one is unable to recover the true SF from the Monte Carlo data sets.
Table \ref{PRH-data-results} shows results for the third case of sampling only, since others give similar results.

\begin{table}
\caption{Results on PRH data: irregular sampling and periodic gaps only, observational errors 
with variance $1\times10^{-7}$, see \S \ref{artificial-prh-section}.}
\label{PRH-data-results}
\begin{center}
  \begin{tabular}{ c c c c} \hline \hline
     $\Delta$ & PRH* 	       & PRH**  	      & Kernels*** \\
   True delay & mean$\pm$std   & mean$\pm$std   & mean$\pm$std   \\ \hline
	34	  & 34.0$\pm$2.3   & 23.2$\pm$21.9  & 33.9$\pm$10.3 \\
	43	  & 43.1$\pm$4.7   & 45.0$\pm$16.2  & 43.4$\pm$2.4\\
	49	  & 50.6$\pm$7.0	 & 47.2$\pm$17.5	& 48.8$\pm$6.9\\
	59	  & 60.0$\pm$5.5	 & 58.9$\pm$21.4	& 59.8$\pm$10.4\\
	66	  & 66.0$\pm$2.5	 & 63.3$\pm$23.2	& 66.6$\pm$9.1 \\
	76	  & 76.7$\pm$3.8	 & 71.1$\pm$22.2	& 75.5$\pm$11.5\\
	99	  & 100.9$\pm$7.2  & 97.2$\pm$21.1	& 103.2$\pm$12.5 \\
    \hline
   \multicolumn{4}{l}{* \ \ \ \ PRH method with SF fixed to true values $c_1$ and $c_2$.}\\
   \multicolumn{4}{l}{** \ \ PRH method with SF estimated from the data, image A.}\\
   \multicolumn{4}{l}{*** Kernels with variable width $k=3$.}
  \end{tabular}
\end{center}
\end{table}

\section{The gravitational lens \object{Q0957+561}: radio observations}\label{0957-section}
In this section we apply the tools developed in this paper to estimate
the time delay for the much studied quasar \object{Q0957+561}.  We use radio
monitoring data at 4~cm and 6~cm wavelengths.  
For the 6~cm data set, we use
the light curve with four points from Spring 1990 removed, as in 
\citep[see][]{Haarsma:1999:TRW}, \footnote {Data
from {\tt http://space.mit.edu/RADIO/papers.html}.  Note that the
6~cm data set has a record not included in the published papers and the
observation on 11th April 1994 is recorded a day earlier in previous
studies.}.  These radio data sets are plotted at the top in
Fig.~\ref{0957-fig}.  Our results are presented in Table~\ref{0957-table}.

To estimate the time delay for this quasar, we use both the fixed
kernel width and variable kernel width approaches outlined in \S
\ref{kernel-method-section}. We employ flux ratios $M=1/1.44$ and
$M=1/1.43$ for the 4~cm and 6~cm data, respectively (the most likely values
given our models).
We tested time delays between $\Delta_{min}=300$ and
$\Delta_{max}=500$, with increments of $1$ day. As in the previous
section, we use a threshold of $0.001$ when regularising matrix
inversion through SVD.  The noise model is assumed to be zero mean
i.i.d. Gaussian with standard deviation of $2$\% of the observed flux
value.

\begin{figure*}[p!]
\centering
\includegraphics[width=17cm]{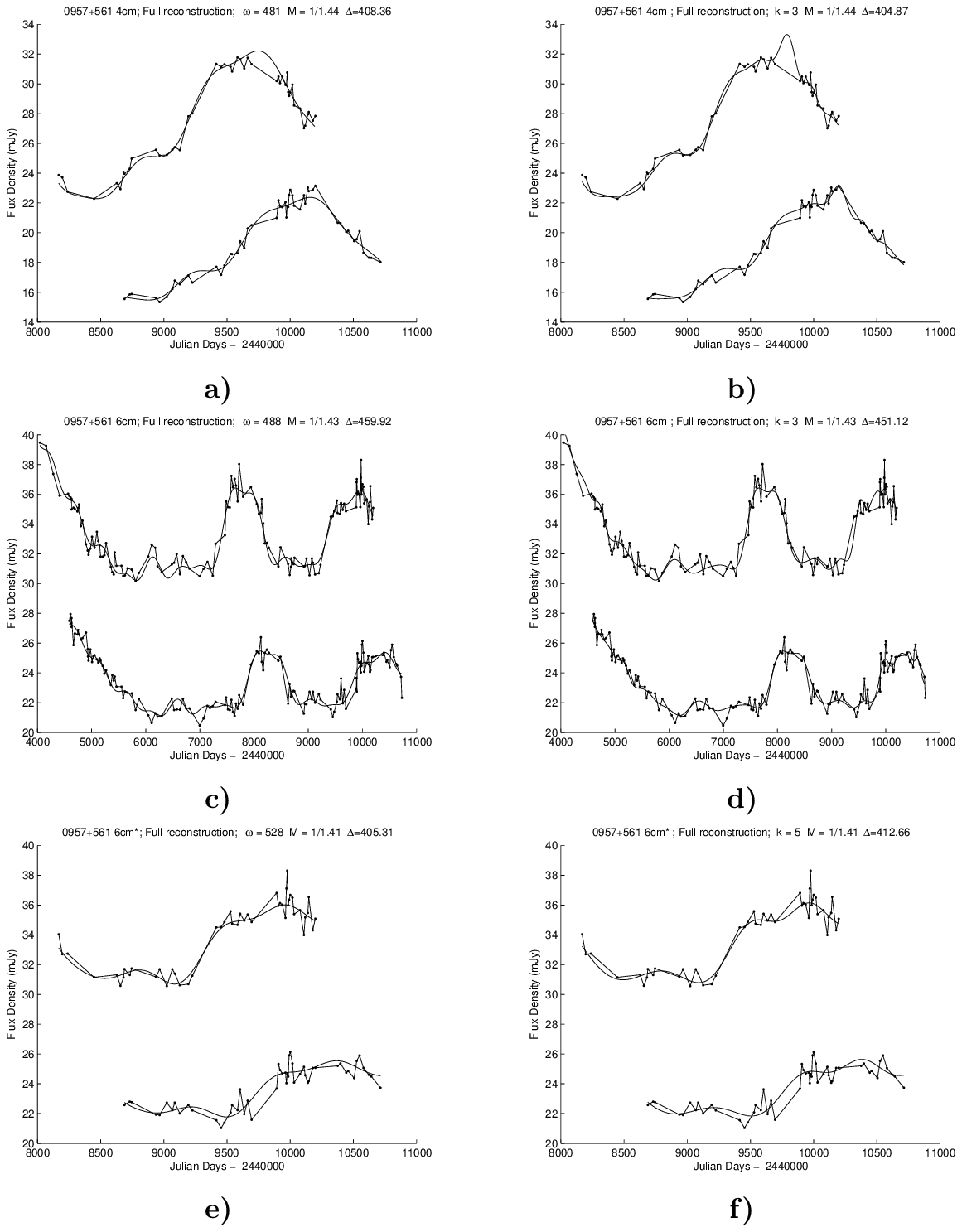}
\caption{Radio observations of gravitationally lensed images A \& B of
\object{Q0957+561}. We show reconstructions with fixed parameters: {\bf a)} 4~cm
with fixed width ($\omega=481$, $\Delta=408.3$), {\bf b)} 4~cm with
variable width ($k=3$, $\Delta=404.8$), {\bf c)} 6~cm with fixed width
($\omega=488$, $\Delta=459.9$), {\bf d)} 6~cm with variable width
($k=3$, $\Delta=451.1$), {\bf e)} 6~cm* with fixed width ($\omega=528$,
$\Delta=405.3$), and {\bf f)} 6~cm* with variable width ($k=5$,
$\Delta=412.6$). Within each plot, at the top is the image A and at
the bottom is image B. The continuous lines are our reconstructed
underlying light curves, $h_{A}(t_{u})$ and $h_{B}(t_{v})$ in Eq.~\ref{fit}.}
\label{0957-reconstruction-fig}
\end{figure*}

For the fixed kernel width technique (\S \ref{CV-section}), 
we use Algorithm \ref{cross-validation} with the following parameters: 
$LowerBound=100$ and $UpperBound=1200$ with increments of $1$ day, 
The selected kernel widths ($\omega$) were $481$ and $488$ days, 
and the estimated time delays were 409 days and 459 days 
for the 4~cm and 6~cm bands, respectively.
To calculate confidence intervals on our time delay estimates,
we performed 500 Monte Carlo simulations 
by adding noise realisations to the observed data.
Confidence intervals were determined as standard deviations
of time delay estimates across the Monte Carlo samples.
We found delays of 408$\pm$10 days and 460$\pm$18 days for 4~cm and 6~cm respectively.
Flux reconstructions with these time delays are shown in
in Figs. \ref{0957-reconstruction-fig} a) and \ref{0957-reconstruction-fig} b).

For the variable kernel width method (\S \ref{k-nearest-section}), 
the number of neighbours $k$ determining local kernel widths
was estimated by Algorithm \ref{cross-validation} 
($\omega$ is replaced by $k$) with 
$LowerBound=1$ and $UpperBound=15$ (increments of $1$).
We obtained $k=3$ for 4~cm, and the estimated time delay was 405 days.
Confidence interval computed on 500 Monte Carlo samples  was 404.8$\pm$11. 
Flux reconstructions with this time delay are shown 
in Fig. \ref{0957-reconstruction-fig} c). 
For 6~cm data, we found $k=3$, and the delay of 450 days.
The 500 Monte Carlo samples gave us a time delay of 451.1$\pm$30 days.
Flux reconstructions are presented in Fig. \ref{0957-reconstruction-fig} d).

Using the PRH method, 
\citet{Haarsma:1999:TRW} report time delays of 397$\pm$12 and 
452$_{-15}^{+14}$ days for the 4~cm and 6~cm data, respectively, 
and 409$\pm$30 on the combined 4+6~cm data set.
They also report
results of the Dispersion spectra method: 
383$_{-19}^{+15}$ and 416$_{-24}^{+22}$ days for the 4~cm and 6~cm
data, respectively, and 395$_{-15}^{+13}$ days on the combined 4+6~cm data set.

There has been a great deal of concern about the difference in time
delay estimates from the two different wavelengths, since
gravitational lensing is achromatic.  Inspired by the results from our
experimentation with artificial data, where the uncertainty of time
delay estimates increases as the gap size increases, we have generated
a new data set, 6~cm*, in order to avoid the effect of the different
gap sizes for different wavelengths.  The 6~cm* data set contains 6~cm
observations sampled only at observation times of the 4~cm dataset. In
other words, we keep a 6~cm observation at time $t$ if there is a 4~cm
observation at the same time $t$.  The 4~cm and 6~cm* data sets both
contain 58 observations.

Time delay estimates obtained by our methods on 500 Monte Carlo
samples based on the 6~cm* data set are presented in Table
\ref{0957-table}.  The `optimal' kernel parameters, $\omega=528$ and
$k=5$, are obtained following the procedure described above.  The
estimated time delays are 405~days and 412~days for the fixed kernel
width and variable kernel width methods, respectively.  The resulting
flux reconstructions are shown in Figs. \ref{0957-reconstruction-fig}
e) and \ref{0957-reconstruction-fig} f).  

On comparing the 4~cm and 6~cm* samples, which are
pairs of the observations at the same epoch and thus have identical
gaps in the time series, we find a consistent value for the estimated
time delay. This exercise indicates that
the disagreement between the 4~cm and 6~cm datasets is largely due to
sampling and systematic errors. 

\begin{table}
\caption{The time delay between \object{Q0957+561} A \& B estimated from
radio ``light'' curves at 4~cm and 6~cm.}
\label{0957-table}
\begin{center}
  \begin{tabular}{ l l l } \hline \hline
   Kernel method: & fixed width    & variable width  \\ \hline
    \textbf{4~cm}   & 408.3$\pm$10 & 404.8$\pm$11 \\
    \textbf{6~cm}   & 459.9$\pm$18 & 451.1$\pm$30 \\
    \textbf{6~cm*}  & 405.3$\pm$29 & 412.6$\pm$35 \\
    \hline
   \multicolumn{3}{l}{\textit{Note}: The time delays are in days.}\\
   \multicolumn{3}{l}{The construction of the {\bf 6~cm*} sample, 
     which contains}\\
\multicolumn{3}{l}{only the 6~cm observations that have a corresponding
4~cm}\\
\multicolumn{3}{l}{observation
at the same epoch, is described in \S\ref{0957-section}.}\\
  \end{tabular}
\end{center}
\end{table}

It is evident that the large variation in the estimates in the values
of time delay, measured in several analyses of the same observed data
sets that we analyse in this paper, is due to the presence of the gaps
in the monitoring at the two wavelengths. Such gaps are unavoidable in
realistic long-term observing programmes, often leading to
unacceptably deviant time delays (in this case, too large by more than
10\%). Several recent analyses have come to this conclusion in various ways
\citep[e.g.][]{Gil-Merino:2002:TDQ,Pindor:2005:DGL,Eigenbrod:2005:TCM}.

\begin{table}
\caption{Results from experiments with artificial data sets.}
\label{methods-table}
\begin{center}
  \begin{tabular}{l l} \hline \hline
  \textbf{Method} & \textbf{Figure} \\ \hline
    Kernel method with fixed width                & Fig. \ref{cv-fig} \\
    Kernel method with variable width             & Fig. \ref{k-nearest-fig} \\
    DCF and LNDCF                                   & Figs. \ref{dcf-fig} and \ref{lndcf-fig}\\
    PRH method, Structure function from A           & Fig. \ref{prh-a-fig} \\
    PRH method, Structure function from B           & Fig. \ref{prh-b-fig} \\
    Dispersion spectra ($D_{1}^{2}$ and $D_{4,2}^{2}$)& Figs. \ref{d1-fig} and \ref{d4-fig}\\ \hline
  \end{tabular}
\end{center}
\end{table}

\section{Conclusions}

We have introduced a new way of measuring the time delay between light
curves of two images of a gravitationally lensed system, based on
generalised linear regression with fixed- and variable-width Gaussian
basis functions (Kernels) (see \S \ref{CV-section} and \S
\ref{k-nearest-section}). On a large set of controlled experiments
using artificially generated data, we compare the accuracy of our methods
with that of other methods used in the
literature for time delay estimation, notably the DCF, LNDCF, PRH and
Dispersion spectra methods (see Table \ref{methods-table}).

Running a controlled set of experiments is essential for a
well-grounded comparison of competing models.  For the artificial
data, unlike in the case of observed fluxes, we have the luxury of
knowing exactly the magnification ratio $M$ and the time delay
$\Delta$; the noise process is also known.  Therefore, we can reliably
measure the bias ($\Delta_{\mu}$, top of
Figs. \ref{cv-fig}--\ref{d4-fig}) and variance ($\Delta_{\sigma}$,
bottom of Figs. \ref{cv-fig}--\ref{d4-fig}) of the time delay
estimates given by the studied methods.  Obviously, one can never
fully measure the bias when estimating the time delay from real
observations.  On the artificial data, our kernel-based methods
presented in this paper came across as the most accurate and stable
methodologies for estimating the time delays between multiple images
of a gravitationally lensed quasar.

Previous attempts at generating similar artificial data have tried to
simulate specific data sets
\citep[see][]{Pijpers:1997:TDT,Burud:2001:ANA}. 
Our artificial data sets contain simulated light 
curves of widely varying (but still realistic) shapes, 
observational gaps and noise levels (these can be made
available on request -- see more plots 
at http://www.cs.bham.ac.uk/$\sim$jcc/artificial). 
At the bottom of Figs. \ref{cv-fig}--\ref{d4-fig}, 
we can observe a general trend  
of increased uncertainty as the gap size 
increases.
The uncertainty is also proportional to the noise level.

Our methods for estimating the time delay introduced in sections
\S \ref{CV-section} and \S \ref{k-nearest-section}, 
give similar results
(see Figs. \ref{cv-fig} \& \ref{k-nearest-fig}),
although the variable kernel width method tends to require
less computational time.

Finally, we have estimated the time delay between of two images of the
quasar \object{Q0957+561} from radio observations at 4~cm and 6~cm.  The time
delay estimates given by our methods are in the range of 405--412 days
(see Table~\ref{0957-table}), which are lower than, but consistent
with, most of the other estimates from the same or similar radio data
sets (see \S\ref{0957-section}).  However, the 6~cm* data set, which
by construction includes only observations that are performed at the
same epoch as those in the 4~cm data set, yields essentially the same
value for the time delay at that obtained from the 4~cm data set
(these can be combined to yield $408\pm 12$ days, the errors being
lower for just the 4~cm data), as opposed to a value of $\sim$450~days
as obtained from the full 6~cm data set, which covers a longer
monitoring period.

We conclude that
such systematic differences between results obtained from observations
at various wavelengths are due to the irregular sampling, and in
particular, due to the presence of large gaps in the monitoring
data. Experiments with simulated data sets like ours help in the
understanding of how the results depends on the sampling, and in
assessing the reliability of the time delays obtained by various
methods.

\begin{acknowledgements}
We thank the anonymous referee for insightful comments that
helped to significantly improve this work. JCCT thanks 
his sponsors, PROMEP and the UASLP. 
\end{acknowledgements}

\bibliography{4652.bbl}

\end{document}